\pdfoutput=1
\documentclass[entropy,article,submit,oneauthor]{Definitions/mdpi}

\firstpage{1}
\pubvolume{1}
\issuenum{1}
\articlenumber{0}
\pubyear{2026}
\copyrightyear{2026}
\datereceived{}
\daterevised{}
\dateaccepted{}
\datepublished{}

\graphicspath{{figure/}}
\newlength{\suppdisplaywidth}
\newenvironment{suppwide}{\begin{adjustwidth}{-\extralength}{0cm}}{\end{adjustwidth}}

\Title{Blind Source Separation Can Distort Behavior and Connectivity Analyses of Calcium Transients}

\Author{Sadiq A. Adedayo $^{1,}$*\orcidA{}}
\AuthorNames{Sadiq A. Adedayo}

\address{%
$^{1}$ \quad UniVie Doctoral School of Computer Science, University of Vienna, Vienna, Austria; sadiq.adedayo@univie.ac.at}

\corres{Correspondence: sadiq.adedayo@univie.ac.at}

\abstract{Denoising is often treated as a technical prelude to calcium transients analysis, but it can redefine the variables used for behavior decoding and causal structure learning. An inspectable motorneuron recording with a visible artefact motivated this study: removing artefact-linked blind source separation (BSS) components also changed trace dynamics outside the targeted frame. We therefore tested whether BSS denoising preserves behavior and causal evidence on already extracted calcium transients. In a synthetic benchmark with known lagged graphs, raw corrupted traces retained graph recovery, whereas component-removing BSS variants collapsed median graph F1 to 0 across estimator families. We then evaluated four BSS methods: FastICA, Infomax, SOBI, and JADE, on four larval zebrafish recordings of v2a reticulospinal neurons (v2a-RSNs) with tail behavior. BSS sometimes improved behavior decoding, but gains depended on fish, method and retained clusters; matched PCA and low-pass controls often matched or exceeded BSS in fold-audited comparisons. Connectivity effects were more consistent: c-GC and c-GC* graphs inferred from BSS-cleaned traces were much denser than raw-trace graphs. BSS should therefore be treated as an intervention on the measured process, not a neutral cleanup step.}

\keyword{blind source separation; calcium imaging; causal structure learning; behavior decoding; data preprocessing}

\begin{document}

\section{Introduction}

Calcium imaging is often used to relate population activity to behavior \cite{Ahrens2012MotorAdaptation, Ahrens2013WholeBrain, Portugues2014WholeBrain, kim2017pan, marques2018structure} and to infer directed interactions among recorded neurons \cite{friston2011functional,gilson2016estimation}. These goals pressure preprocessing. Fluorescence traces are shaped by measurement error, indicator kinetics, sampling, motion, source overlap, neuropil contamination and extraction choices \cite{carroll2006measurement, GrienbergerKonnerth2012Calcium, Chen2013GCaMP, pnevmatikakis2017normcorre, Pnevmatikakis2016Calcium, giovannucci2019caiman,pnevmatikakis2019analysis,kim2022fluorescence,vanwalleghem2021calcium}. Post extraction denoising changes trace appearance and the stochastic process supplied to behavior decoders and causal structure learning (CSL) algorithms.

Causal analysis assumes that measured variables and their histories preserve dependencies needed for inference \cite{spirtes2000causation, pearl2009causality, peters2017elements, runge2023causal}. Filtering, downsampling, temporal aggregation and windowed inference can change Granger causality, transfer entropy and lagged conditional independence tests \cite{florin2010filtering, barnett2011filtering, Seth3293, seth2013fmri, weber2017transfer, luo2013spatiotemporal, zhou2014sampling, barnett2017subsampling, gong2017temporal, hyttinen2017subsampled, plis2015rateagnostic}. Denoising should therefore be validated as an intervention on the measured process.

We focus on blind source separation (BSS) after trace extraction. FastICA, Infomax, SOBI and JADE can expose latent components and aid artefact separation \cite{Hyvarinen1999FastICA, BellSejnowski1995Infomax, belouchrani1997sobi, CardosoSouloumiac1993JADE, HyvarinenOja2000ICA, BrownYamadaSejnowski2001NeuralICA, VigarioOja2008Neuroinformatics, Mukamel2009CaImaging}. But component rejection learns a basis, deletes coordinates and reconstructs traces. Discarded components may contain behavior-relevant activity, while retained neural-looking components may still contain artefact. A useful cleaned representation must preserve trace geometry, behavior information and temporal state structure.

The starting point was a small motorneuron recording with a visible artefact, previously used in Granger causality analyses of calcium traces \cite{chen2023granger, fallani2014hierarchy}. Applying BSS to that recording targeted the obvious artefact but also changed trace dynamics outside the defective frame, motivating the main question of whether component rejection preserves the temporal evidence used for behavior decoding and CSL. We therefore built an automated BSS pipeline for higher dimensional calcium recordings in which artefacts are not identified in advance. The empirical analysis focuses on four v2a reticulospinal neuron (v2a-RSN) recordings with tail behavior \cite{carbo2022functional}, and a synthetic calcium trace benchmark with known lagged graphs, injected artefacts and matched BSS reconstructions validates the evaluation pipeline under ground truth.

The main contribution is a validation framework for BSS denoising before causal analysis. It combines spectral component clustering, controlled reconstruction, matched PCA and low pass controls, behavior scorecards, fold local leakage audits, latent state diagnostics and graph sensitivity. This framing follows work showing that trace-domain and deconvolved representations can suit different downstream analyses \cite{shen2021deconvolve}. The goal is not to rank BSS methods, but to test whether a preprocessing step that helps one readout distorts evidence needed for another.

\section{Materials and Methods}
\label{sec:methods}

\subsection{Data representation and evaluation regimes}

Let \(D\in\mathbb{R}^{N\times T}\) denote the extracted fluorescence or spike rate trace matrix, with \(N\) neurons or regions of interest and \(T\) calcium frames. For decomposition and reconstruction we use \(X=D^\top\in\mathbb{R}^{T\times N}\), where rows index time and columns index neurons. A trace representation is written \(X^{(m)}\), with \(m=0\) denoting raw traces and \(m>0\) denoting a BSS cleaned reconstruction or matched control.

Two regimes are retained. The descriptive regime fits decomposition and spectral feature clustering to a full recording before blocked decoder evaluation. It supports case study inspection but is transductive. The audited regime fits learned operations inside each fold. It places BSS, clustering, selection, baselines, decoders and causal state diagnostics inside the training boundary for each held out block.

\subsection{Causal assumptions}
\label{sec:causal_assumptions}

The causal language is grounded in the standard observational CSL setting, where graphical claims depend on causal Markov structure, faithfulness, finite timeseries memory and causal sufficiency when a graph is given a causal interpretation \cite{spirtes2000causation, pearl2009causality, peters2017elements, runge2023causal}. These assumptions are used as diagnostic criteria, not guarantees: observed history must approximate the effective state, conditional independences should reflect graph separation rather than cancellations or preprocessing artefacts, and sampled variables should block relevant common causes. The v2a-RSN recordings sample only part of a sensorimotor circuit, so unobserved neurons, sensory drive, motion and behavior feedback may confound the traces; lagged conditioning helps but does not prove sufficiency. The reported connectivity matrices are therefore sensitivity analyses of the measured process, not validated effectomes.

\subsection{Synthetic known-graph benchmark}
\label{sec:methods_synthetic_benchmark}

The simulation benchmark generated latent vector autoregressive activity with a
known sparse lagged graph, calcium-like observation dynamics, additive artefact
processes and behavior targets derived from delayed latent parents. Eight
scenarios varied artefact strength and stress conditions. Each scenario was run
for 30 random seeds. For each replicate, the runner compared clean traces, raw
corrupted traces, artefact-subtraction oracles, all-component BSS
reconstructions, component-removing BSS reconstructions, PCA, random-subspace
and low-pass controls.

Graph recovery was scored against the known directed lagged graph after
collapsing estimator outputs to off-diagonal directed adjacency matrices. The
reported metrics include precision, recall, F1, specificity, false positive
rate, derived false discovery rate, Matthews correlation coefficient,
structural Hamming distance, edge-density bias and direction reversals. The
c-GC and c-GC* simulation reruns used the fast c-GC backend with
\(n_{\mathrm{perm}}=1000\), \(\alpha=0.01\), \(\beta=0.005\), no graph
correction and contemporaneous edges excluded from the primary graph. PCMCI+
and JPCMCI+ used their native Tigramite graph decisions. Summary tables report
medians over the validated aggregate outputs.

\subsection{Fold local execution}
\label{sec:strict_evaluation}

The audited configuration uses five contiguous held out test blocks with a 14-frame exclusion gap around each block. Training frames may occur before and after a test block, but no training example shares a raw frame with a held out history. For fold \(f\), \(I_f^{\mathrm{train}}\) and \(I_f^{\mathrm{test}}\) denote train and test indices, and \(G_f\) denotes the excluded gap. Here \(\Phi_t\) is the input feature vector available at time \(t\), \(y_t\) is the corresponding target, \(\mathcal{H}\) is the candidate class of learned maps, \(\ell\) is the training loss, and \(Q\) is the held out scoring functional. For any learned map \(h\),
\begin{align}
    h_f
    &=
    \operatorname*{arg\,min}_{h\in\mathcal{H}}
    \sum_{t\in I_f^{\mathrm{train}}}
    \ell\!\left(y_t,h(\Phi_t)\right),
    \nonumber\\
    \operatorname{score}_f
    &=
    Q\!\left(\{y_t,h_f(\Phi_t)\}_{t\in I_f^{\mathrm{test}}}\right).
    \label{eq:fold_local_rule}
\end{align}
\(h_f\) is the fold-specific map chosen using only training indices, and
\(\operatorname{score}_f\) is computed only from held out predictions in that
fold.
All decomposition, clustering, component selection, baseline fitting, decoder fitting and causal state fitting steps obey this rule.

\subsection{BSS decomposition and reconstruction}

The four BSS methods are FastICA, Infomax, SOBI and JADE \cite{Hyvarinen1999FastICA,BellSejnowski1995Infomax,belouchrani1997sobi,CardosoSouloumiac1993JADE,HyvarinenOja2000ICA,BrownYamadaSejnowski2001NeuralICA,VigarioOja2008Neuroinformatics}. For a centered training matrix \(\tilde{X}\), decomposition gives source time courses \(S\in\mathbb{R}^{T\times K}\) and loadings \(A\in\mathbb{R}^{N\times K}\). Full reconstruction is
\begin{equation}
    \hat{X}_{\mathrm{full}} = S A^\top + \mu .
    \label{eq:methods_full_reconstruction}
\end{equation}
\(\hat{X}_{\mathrm{full}}\) is the all-component reconstruction, and \(\mu\)
is the neuron-wise mean vector, broadcast across time.
For retained component set \(R\), the cleaned reconstruction is
\begin{equation}
    \hat{X}_{R} = S_{:,R} A_{:,R}^\top + \mu .
    \label{eq:methods_clean_reconstruction}
\end{equation}
\(S_{:,R}\) and \(A_{:,R}\) select the columns associated with retained
components in \(R\), so \(\hat{X}_{R}\) is the reconstruction after component
selection.
Infomax, SOBI and JADE use a common PCA whitening operator before rotation. The Supplementary Methods give the shared joint diagonalization routine used by SOBI and JADE.

\subsection{Component features and selection}

Components are scored with temporal, spectral, loading and optional behavior features. Welch power spectra provide power ratios, entropy and narrowband dominance \cite{Welch1967PSD}. Components are clustered with \(k\)-means in spectral feature space \cite{Lloyd1982KMeans}. Candidate reconstructions retain ordered cluster sets, and the descriptive v2a-RSN sweeps vary retained cluster count. The Supplementary Methods define the feature table and reconstruction impact metrics.

\subsection{Behavior targets and BPI}

High-rate tail angle samples are binned to calcium frames. Tail vigor is the root mean square of first difference tail velocity within each frame bin. Bout state is thresholded from vigor. The Supplementary Methods give the formal target construction.

For behavior target \(q\), representation \(m\) and shift \(s\), the target score is
\begin{equation}
    M_{m,q,s}
    =
    Q\!\left(y^{(q)}_{t+s}, \hat{y}^{(q,m)}_{t+s}\right),
    \label{eq:target_metric}
\end{equation}
Here \(M_{m,q,s}\) is the held-out target score, \(y^{(q)}_{t+s}\) is the
observed shifted target, \(\hat{y}^{(q,m)}_{t+s}\) is the corresponding
prediction, and \(Q\) is Pearson correlation for continuous targets and
balanced accuracy for bout state. BPI compares cleaned and raw decoding:
\begin{equation}
    \mathrm{BPI}_{m}
    =
    100
    \times
    \frac{1}{|\mathcal{Q}|}
    \sum_{q\in \mathcal{Q}}
    \frac{M_{m,q,s_q}}{M_{0,q,s_q}} .
    \label{eq:methods_bpi}
\end{equation}
\(\mathcal{Q}\) is the set of behavior targets included in the scorecard,
\(s_q\) is the selected shift for target \(q\), and \(M_{0,q,s_q}\) is the raw
trace reference score. Values above 100 indicate stronger behavior decoding
than raw traces under this scorecard.

\subsection{Causal state diagnostics}

The causal state diagnostic asks whether cleaned traces preserve latent dynamics as well as behavior information. For representation \(m\), histories are
\begin{equation}
    H^{(m)}_t =
    \left[
        X^{(m)}_{t-w+1,:},
        \ldots,
        X^{(m)}_{t,:}
    \right].
    \label{eq:history_windows}
\end{equation}
\(w\) is the history length in frames, and the bracket denotes concatenation of
all neuron vectors in the window ending at time \(t\).
A fold local PCA encoder maps histories to a 3-dimensional state
\(z^{(m)}_t\), and a residual transition model predicts
\begin{equation}
    \hat{z}^{(m)}_{t+1}=z^{(m)}_t+f(z^{(m)}_t).
    \label{eq:transition_model}
\end{equation}
\(z^{(m)}_t\) is the encoded low-dimensional state, \(f\) is the fitted
transition residual map, and \(\hat{z}^{(m)}_{t+1}\) is the predicted next
state.
Linear ridge transitions are the main diagnostic, following the use of simple state-prediction probes as audit baselines \cite{kumar2023bundle,grosseWentrup2024ncmcm}; shallow tanh MLP transitions are nonlinear sensitivity checks. Extra history tests ask whether older history improves held out prediction after conditioning on \(z_t\). Transition residuals are also tested against behavior history, artefact probes, rejected component scores and raw trace directions. A cleaned variant is treated as state preserving only when behavior scores, dynamic error, extra history dependence and artefact probe sensitivity are all acceptable relative to raw traces and matched controls.

\subsection{Baselines, uncertainty and aggregation}

The audited evaluation compares BSS variants with raw traces, rank matched PCA \cite{TippingBishop1999PCA}, seeded random orthogonal subspaces, energy matched PCA and causal one sided low pass filters. All learned baselines are fit inside the same training boundary as BSS. Supplementary Table S2 lists ranks, seeds, low pass cutoffs, target shifts and selection rules.

Prediction level uncertainty is computed within each recording. Regression effects are paired differences between raw and variant squared errors; classification effects are paired differences between variant and raw correctness. Fixed nonoverlapping 60-frame blocks are resampled for 2,000 bootstrap replicates and sign flipped for 5,000 permutation replicates, with 95\% percentile intervals. Blocked evaluation follows time-respecting decoder practice and avoids random frame splits for autocorrelated data \cite{Varoquaux2017DecoderCV,Vabalas2019LimitedSample,CawleyTalbot2010SelectionBias}. Aggregation keeps recordings and fish, not cross validation folds or frames, as the biological inferential units.

\subsection{Connectivity estimation}

Motorneuron traces were passed through c-GC, c-GC*, PCMCI+ and JPCMCI+ variants with fixed lag \(\tau=2\) \citep{adedayo2025cgc,runge2019detecting,gunther2023jpcmci}. v2a-RSN traces were passed through c-GC and c-GC* using \(\tau=2\) and \(n_{\mathrm{pasts}}=4\). The lag \(\tau=2\) was an assumed value corresponding to approximately one-third of the acquisition frequency, and \(n_{\mathrm{pasts}}=4\) was selected as a sufficient conditioning depth for \(\tau=2\); see Adedayo et al. \citep{adedayo2026markovianityDepth}. v2a-RSN PCMCI+ and JPCMCI+ variants were omitted because of runtime and computational cost. Connectivity matrices are reported as sensitivity outputs because neither dataset provides an edge level ground truth graph for all reported estimators.

\section{Results}

\subsection{Known-graph simulations separate cleaning from graph recovery}
\label{sec:nm_simulation_results}

The synthetic benchmark supplies an edge-level ground truth that the empirical
recordings lack. It generated 8 artefact and stress scenarios across 30 seeds
for each BSS split, giving 240 validated replicates for each of the four
BSS methods. The c-GC variants were rerun using \(n_{\mathrm{perm}}=1000\),
\(\alpha=0.01\), and \(\beta=0.005\), while PCMCI+ and JPCMCI+ were scored from
their native Tigramite graph decisions. Figure~\ref{fig:nm_simulation_construction}
shows the two simulated ingredients most relevant to BSS evaluation: additive
artefact and delayed behavior targets.

\begin{adjustwidth}{-\extralength}{0cm}
    \begin{minipage}{\fulllength}
        \centering
        \begin{minipage}[t]{0.49\linewidth}
            \centering
            \includegraphics[width=\linewidth]{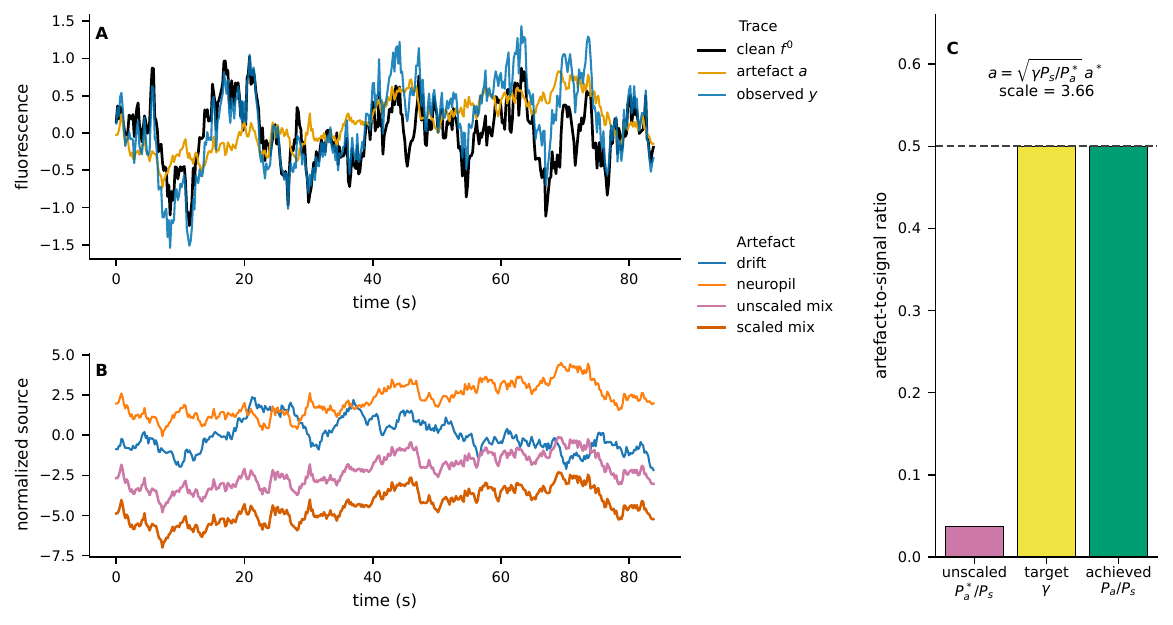}
            \par\smallskip
            {\footnotesize\textbf{a.} Artefact construction and scaling.\par}
        \end{minipage}
        \hfill
        \begin{minipage}[t]{0.49\linewidth}
            \centering
            \includegraphics[width=\linewidth]{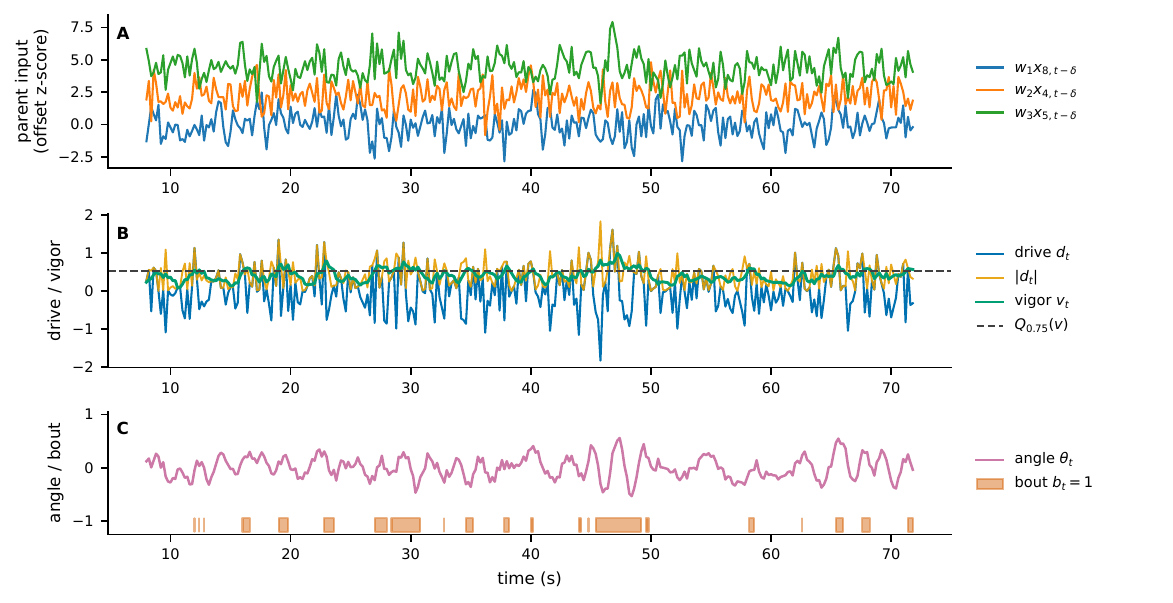}
            \par\smallskip
            {\footnotesize\textbf{b.} Behavior targets from delayed latent parents.\par}
        \end{minipage}
        \captionof{figure}{Synthetic benchmark construction. (a) Artefact scaling for
        one replicate. (b) Delayed latent-parent behavior targets.}
        \label{fig:nm_simulation_construction}
    \end{minipage}
\end{adjustwidth}
\vspace{0.6\baselineskip}

Raw corrupted traces retained recoverable graph information. Median raw graph
F1 was 0.585 for c-GC, 0.500 for c-GC* and 0.658 for J-/PCMCI+
(Table~\ref{tab:nm_simulation_graph_failure_modes}). Clean and artefact-oracle
references were higher, with median F1 values of 0.750, 0.714 and 0.857 for
the same estimator families. The main negative result appears after actual
component removal. Across all \texttt{cluster\_keep\_top} BSS variants, median
graph F1 was 0 for all estimator families. Even when the best
component-removing BSS variant was selected within each scenario and BSS
method, the median F1 stayed at 0 for c-GC and c-GC*, and reached only 0.077
for J-/PCMCI+.

\begin{adjustwidth}{-\extralength}{0cm}
    \centering
    \begin{minipage}{\fulllength}
        \centering
        \begin{minipage}[t]{\linewidth}
            \centering
            \includegraphics[width=0.9\linewidth]{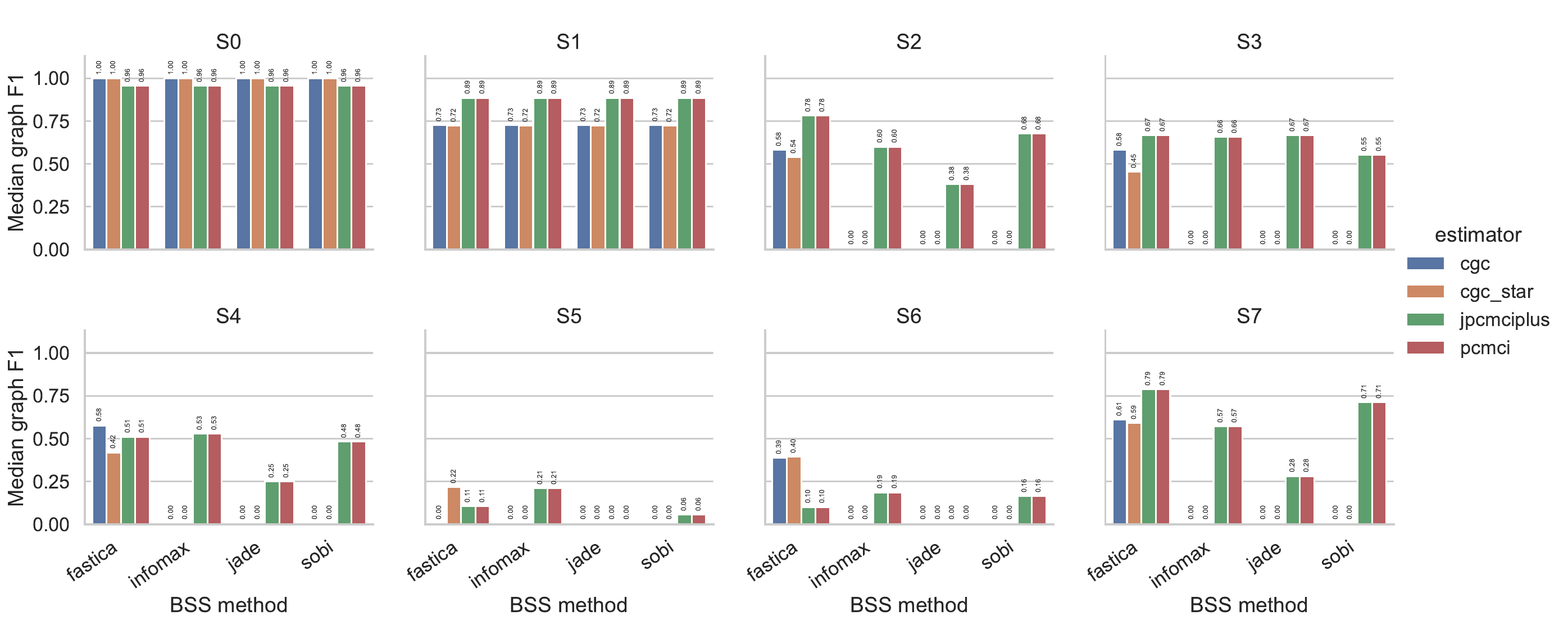}
            \par\smallskip
            {\footnotesize\textbf{a.} Upper-envelope graph F1 by scenario.\par}
        \end{minipage}
        \par\vspace{0.3em}
        \begin{minipage}[t]{\linewidth}
            \centering
            \includegraphics[width=0.9\linewidth]{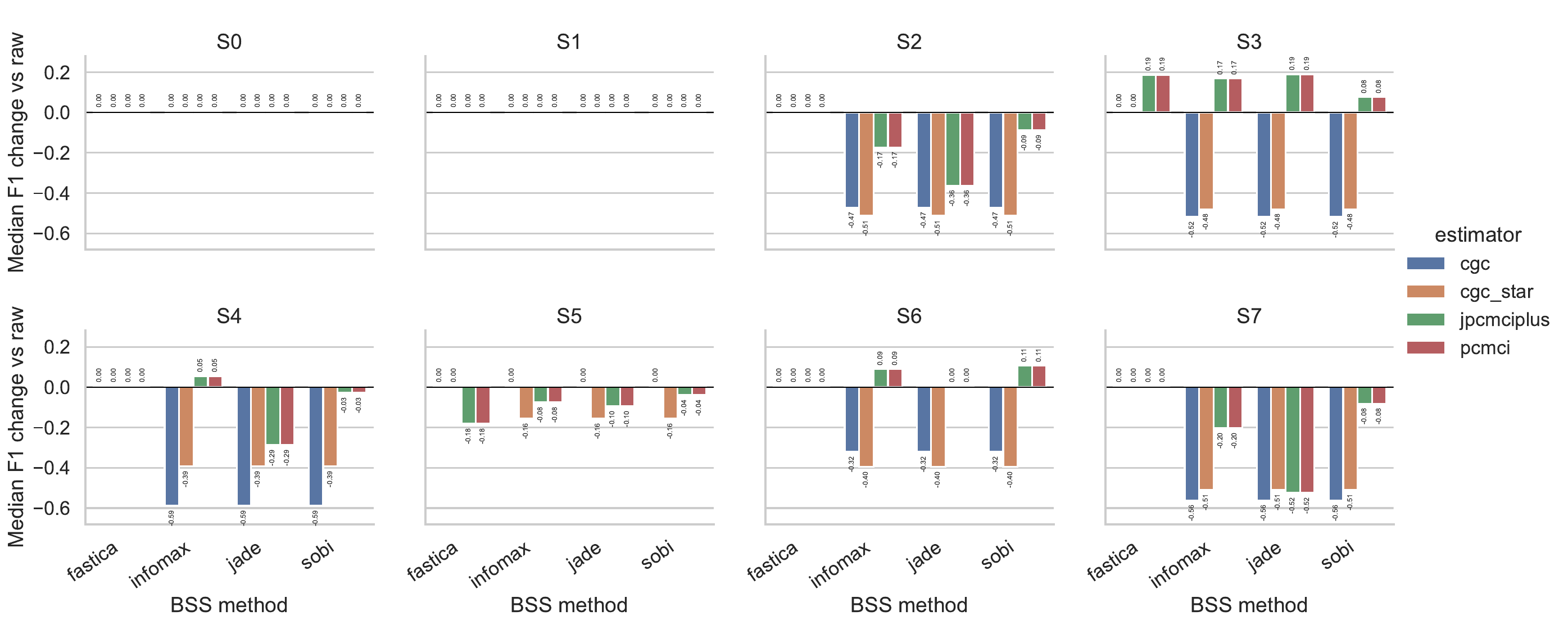}
            \par\smallskip
            {\footnotesize\textbf{b.} Graph F1 change relative to raw traces.\par}
        \end{minipage}
        \captionof{figure}{Synthetic graph-recovery diagnostics. (a) Scenario-level upper
        envelope by BSS method and estimator; plotted best variants include
        all-component and oracle-selection outputs and should be read
        diagnostically. (b) Paired graph F1 change relative to raw traces;
        negative values indicate reduced graph recovery.}
        \label{fig:nm_simulation_graph_recovery_diagnostics}
    \end{minipage}
\end{adjustwidth}
\vspace{0.6\baselineskip}

The degradation was not a single false positive effect. Raw c-GC and c-GC* 
produced more erroneous discoveries than J-/PCMCI+, with median
false discovery rates (FDR) of 0.391 and 0.532 compared with 0.200. In contrast,
component-removing BSS variants usually returned empty or nearly empty graphs:
median true positives and false positives were 0, and median false
negatives were 17--18. The simulation therefore separates two errors. Raw c-GC
variants were relatively liberal, whereas BSS component rejection mostly
destroyed graph-relevant signal.

Figure~\ref{fig:nm_simulation_graph_recovery_diagnostics} summarizes the graph
recovery diagnostics. The upper panel shows the scenario-level upper envelope
when every BSS output family is allowed, including all-component and
oracle-selection variants. This is a diagnostic upper envelope, not a deployable
cleaning result. It often recovers the raw-like all-component reconstruction
rather than a component-removed trace. The lower panel shows the same
limitation: the best available BSS setting rarely improves on raw traces and
often reduces graph recovery in artefact and stress scenarios.
Trace preservation and graph recovery also separated
(Supplementary Figure S4): raw
traces had a median trace correlation of 0.83 against clean fluorescence,
whereas the component-removing BSS variants had a median trace correlation of
0.60. Thus, deleting components often removed graph-relevant signal along with
artefact.

\begin{adjustwidth}{-\extralength}{0cm}
    \centering
    \begin{minipage}[t]{0.57\fulllength}
        \vspace{0pt}
        \centering
        \captionof{table}{Synthetic graph-recovery failure modes. Entries are medians over the validated aggregate outputs. FDR is \(1-\)precision when at least one edge is predicted; empty median graphs have zero false discoveries but zero recall.}
        \label{tab:nm_simulation_graph_failure_modes}
        \footnotesize
        \setlength{\tabcolsep}{2.4pt}
        \renewcommand{\arraystretch}{0.9}
        \begin{tabular}{llrrrrrr}
        \toprule
        \shortstack{Trace\\class} & Learner & F1 & P & R & FDR & FPR & SHD \\
        \midrule
        \multirow{3}{*}{\shortstack{Clean or\\oracle}} & c-GC & 0.750 & 0.833 & 0.686 & 0.125 & 0.027 & 8.0 \\
         & c-GC* & 0.714 & 0.811 & 0.667 & 0.129 & 0.027 & 9.0 \\
         & J-/PCMCI+ & 0.857 & 0.875 & 0.841 & 0.125 & 0.027 & 5.0 \\
        \midrule
        \multirow{3}{*}{Raw} & c-GC & 0.585 & 0.580 & 0.622 & 0.391 & 0.068 & 15.0 \\
         & c-GC* & 0.500 & 0.468 & 0.575 & 0.532 & 0.117 & 19.5 \\
         & J-/PCMCI+ & 0.658 & 0.786 & 0.558 & 0.200 & 0.028 & 11.0 \\
        \midrule
        \multirow{3}{*}{\shortstack{BSS\\removed}} & c-GC & 0.000 & 0.000 & 0.000 & 0.000 & 0.000 & 18.0 \\
         & c-GC* & 0.000 & 0.000 & 0.000 & 0.000 & 0.000 & 18.0 \\
         & J-/PCMCI+ & 0.000 & 0.000 & 0.000 & 0.000 & 0.000 & 18.0 \\
        \bottomrule
        \end{tabular}
    \end{minipage}
    \hfill
    \begin{minipage}[t]{0.40\fulllength}
        \vspace{0pt}
        \centering
        \captionof{table}{Post hoc upper envelope whole recording BPI variant.}
        \label{tab:v2a_transductive_summary}
        \footnotesize
        \setlength{\tabcolsep}{2.0pt}
        \renewcommand{\arraystretch}{0.95}
        \begin{tabular}{rrrlrrr}
        \toprule
        Fish & \shortstack{Vigor\\\(r\)} & \shortstack{Bout\\BA} & \shortstack{Best\\variant} & BPI & \shortstack{Trace\\\(r\)} & NRMSE \\
        \midrule
        1 & 0.597 & 0.736 & SOBI-k2    & 104.0 & 0.719 & 0.662 \\
        2 & 0.038 & 0.495 & Infomax-k2 & 418.4 & 0.847 & 0.536 \\
        3 & 0.365 & NA & Infomax-k1 & 128.2 & 0.845 & 0.558 \\
        4 & 0.251 & 0.601 & Infomax-k2 & 123.6 & 0.848 & 0.528 \\
        \bottomrule
        \end{tabular}
    \end{minipage}
\end{adjustwidth}

\subsection{Whole recording v2a-RSN sweeps reveal behavior and trace tradeoffs}
\label{sec:v2aBSSRes}

The larger test used four v2a-RSN fluorescence recordings with aligned tail behavior. Components were clustered by spectral features, retained cluster count was varied and raw and cleaned traces were tested against tail vigor and bout state. These whole recording sweeps are descriptive because decomposition and cluster selection used the complete recording before blocked decoder evaluation.

The Infomax example in Fig.~\ref{fig:v2a_infomax_cluster_diagnostics} shows how independent components were grouped in spectral feature space. The right panel reports the mean log power spectral density of each cluster, which was used to order clusters for incremental reconstruction.

The descriptive behavior preservation index (BPI) varied strongly across fish, method and retained clusters (Table~\ref{tab:v2a_transductive_summary}; Fig.~\ref{fig:v2a_bpi_diagnostics}). BPI is normalized so that raw traces equal 100. Values above 100 mean stronger behavior decoding under this scorecard, not causal validity or minimal trace distortion. Bout state decoding was not finite for \texttt{fish-3}, so that recording is interpreted through tail vigor decoding and trace preservation metrics.

\subsection{Fold local audits weaken the case for BSS as a reliable denoiser}
\label{sec:v2a_strict_results}

The leakage-audited runner evaluates learned operations inside each fold. It includes all four BSS methods, matched PCA and low-pass controls, behavior decoders and causal-state diagnostics. The full fold-local grid was reported for \texttt{fish-1}, \texttt{fish-2} and \texttt{fish-4}; \texttt{fish-3} was omitted because its 412-variable state space makes the full grid computationally too expensive. Audit tables for the reported runs show no held-out frames inside fitted operations. The remaining boundary issue is input-level preprocessing: nonfinite imputation and centered behavior smoothing occur before fold construction.

The audited results do not support a simple BSS success story. The best BSS variant improved several raw behavior readouts, but matched controls often matched or exceeded BSS (Table~\ref{tab:v2a_strict_controls}; Fig.~\ref{fig:v2a_strict_three_fish_audit_summary}). Best-of-grid entries are selected only from candidates present in all five folds.

\begin{adjustwidth}{-\extralength}{0cm}
\centering
\captionof{table}{Three-recording leakage audited v2a-RSN behavior and state diagnostics.}
\label{tab:v2a_strict_controls}
\setlength{\tabcolsep}{3.4pt}
\footnotesize
\begin{tabular}{rrrrrrrrrr}
\toprule
\shortstack{Fish} &
\shortstack{Raw\\vigor} &
\shortstack{BSS\\vigor} &
\shortstack{Ctrl.\\vigor} &
\shortstack{Raw\\bout} &
\shortstack{BSS\\bout} &
\shortstack{Ctrl.\\bout} &
\shortstack{Raw\\dyn.} &
\shortstack{BSS\\dyn.} &
\shortstack{Ctrl.\\dyn.} \\
\midrule
1 & 0.594 & 0.607 & \textbf{0.624} & 0.736 & 0.761 & \textbf{0.770} & 0.0078 & 0.0072 & \textbf{0.0001} \\
2 & 0.033 & \textbf{0.149} & 0.108 & 0.500 & \textbf{0.560} & 0.533 & 0.0041 & 0.0027 & \textbf{0.0018} \\
4 & 0.247 & 0.341 & \textbf{0.354} & 0.602 & 0.624 & \textbf{0.641} & 0.0085 & 0.0068 & \textbf{0.0016} \\
\bottomrule
\end{tabular}
\end{adjustwidth}

Causal-state diagnostics gave the same warning. Some BSS variants improved normalized dynamic MSE or behavior-head scores relative to raw traces, but the best normalized dynamic MSE in all three audited recordings came from the 0.1 Hz low-pass control. A heavily smoothed representation can make one-step latent prediction easier while discarding information needed elsewhere. Residual artefact null \(Z\) scores and forward--reverse gaps were not estimated; we therefore withhold a state-preservation claim for BSS.

\begin{adjustwidth}{-\extralength}{0cm}
    \centering
    \begin{minipage}[t]{0.49\fulllength}
        \vspace{0pt}
        \centering
        \includegraphics[width=\linewidth]{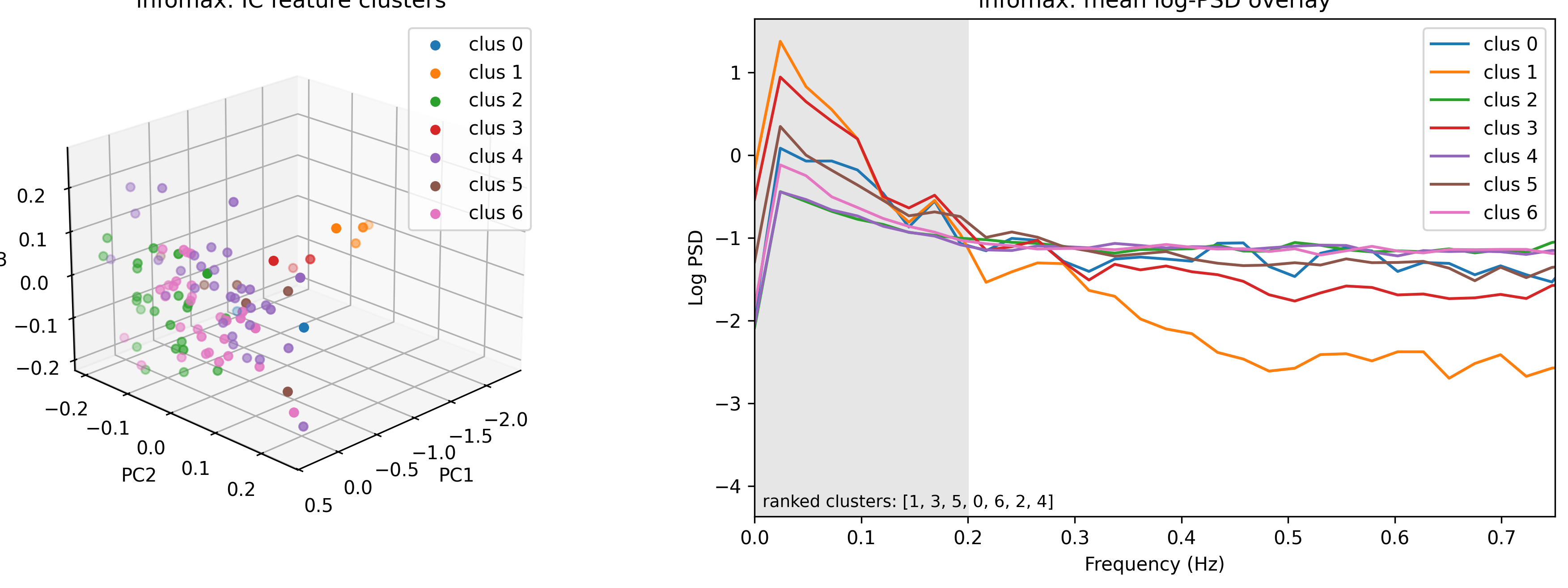}
        \captionof{figure}{Infomax IC cluster diagnostics for the \texttt{fish-1} v2a-RSN fluorescence recording. Left, IC feature clusters; right, cluster mean log power spectral density.}
        \label{fig:v2a_infomax_cluster_diagnostics}
    \end{minipage}
    \hfill
    \begin{minipage}[t]{0.49\fulllength}
        \vspace{0pt}
        \centering
        \captionof{table}{v2a-RSNs c-GC family CSL sensitivity to BSS preprocessing. Entries are off diagonal directed edge densities (\%).}
        \label{tab:v2a_csl_density}
        \footnotesize
        \setlength{\tabcolsep}{2.2pt}
        \renewcommand{\arraystretch}{0.9}
        \begin{tabular}{rrlrrrrr}
        \toprule
        Fish & N & \shortstack{CSL\\method} & Raw & \shortstack{Fast\\ICA} & Infomax & SOBI & JADE \\
        \midrule
        1 & 100 & c-GC  & 4.8 & 86.6 & 84.3 & 84.7 & 87.5 \\
          &     & c-GC* & 4.5 & 87.0 & 86.2 & 84.6 & 87.8 \\
        \midrule
        2 &  92 & c-GC  & 6.7 & 51.9 & 52.9 & 49.3 & 54.0 \\
          &     & c-GC* & 6.6 & 51.5 & 53.0 & 49.3 & 54.5 \\
        \midrule
        4 & 165 & c-GC  & 5.5 & 57.7 & 62.2 & 58.1 & 67.7 \\
          &     & c-GC* & 5.2 & 57.6 & 62.1 & 58.1 & 67.9 \\
        \bottomrule
        \end{tabular}
    \end{minipage}
\end{adjustwidth}
\vspace{0.4\baselineskip}

Figure~\ref{fig:v2a_strict_three_fish_audit_summary} condenses the audited grid
to three representation classes: raw traces, the best BSS variant, and the best
matched control. This avoids plotting the full candidate list while preserving
the main contrast: BSS can help behavior readouts, but the control remains
competitive and gives the lowest dynamic MSE.

\begin{adjustwidth}{-\extralength}{0cm}
    \begin{minipage}{\fulllength}
    \centering
    \begin{minipage}[t]{0.49\linewidth}
        \centering
        \includegraphics[width=0.86\linewidth]{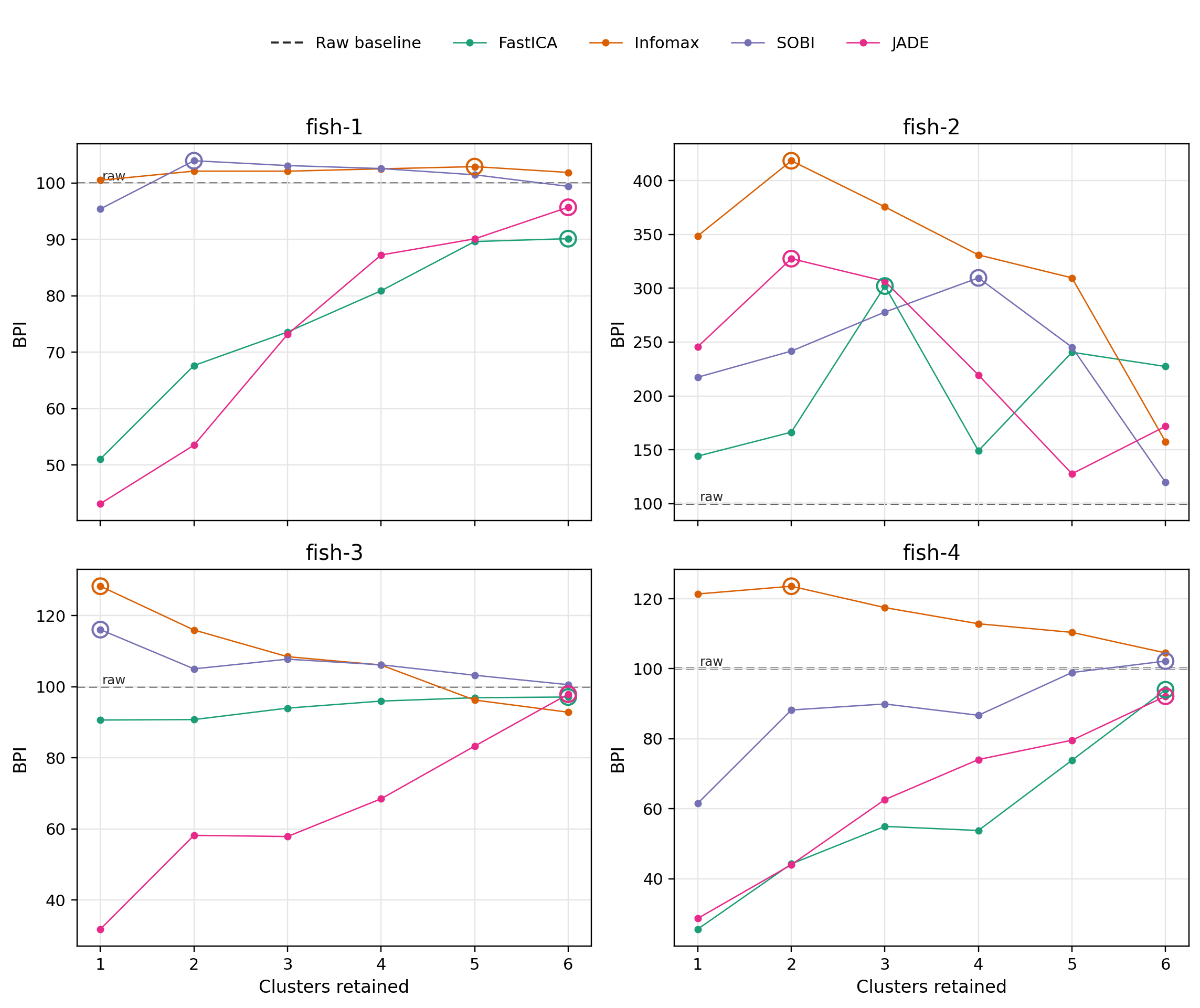}
        \par\smallskip
        {\footnotesize\textbf{a.} BPI as retained clusters are added.\par}
        \label{fig:v2a_cluster_bpi_progression}
    \end{minipage}
    \hfill
    \begin{minipage}[t]{0.49\linewidth}
        \centering
        \includegraphics[width=0.86\linewidth]{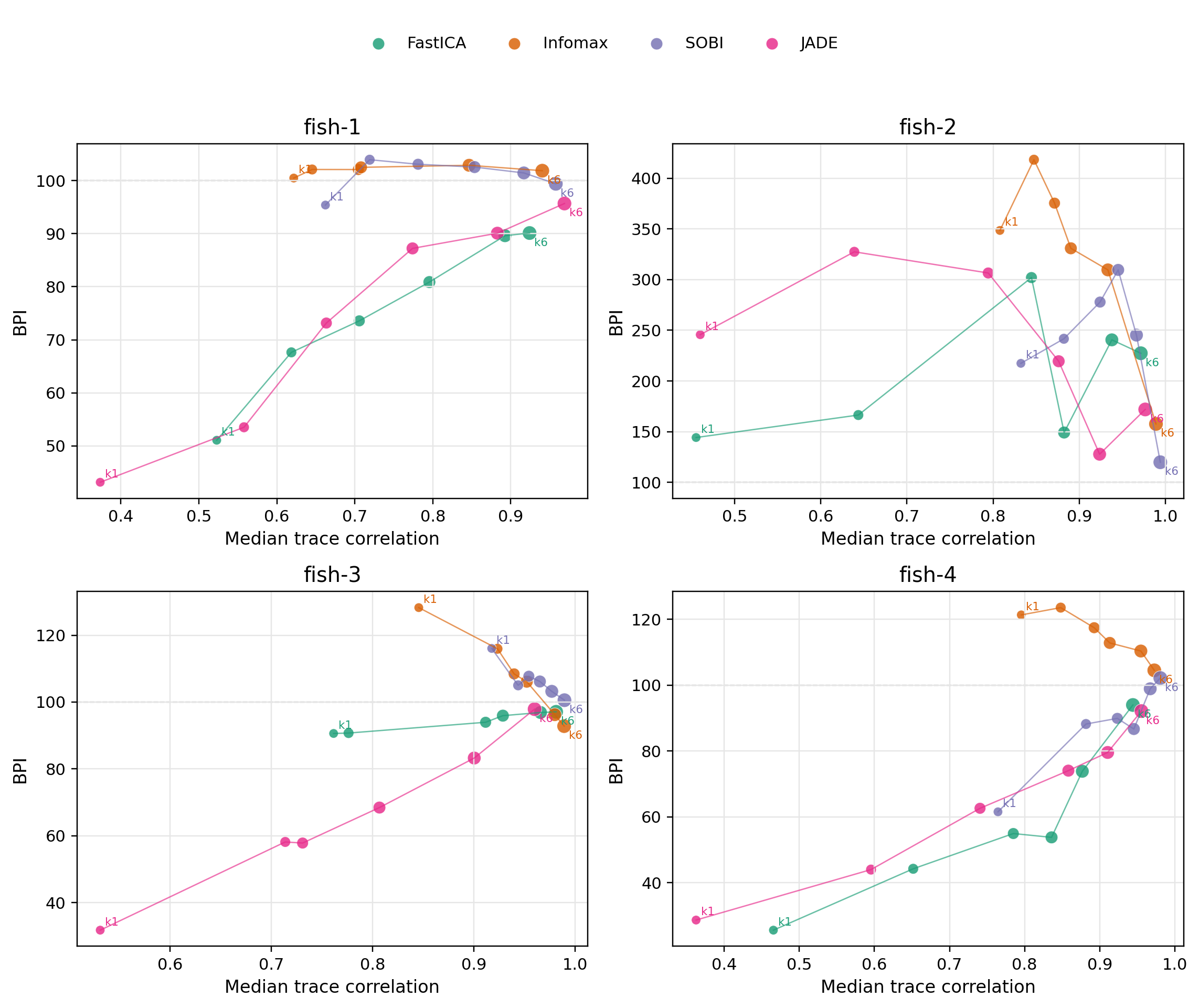}
        \par\smallskip
        {\footnotesize\textbf{b.} BPI against trace preservation.\par}
        \label{fig:v2a_bpi_trace_tradeoff}
    \end{minipage}
    \captionof{figure}{Whole recording BPI diagnostics. (a) Values are normalized to raw traces as retained spectral clusters are added. (b) Retaining more clusters generally increases trace preservation, but behavior decoding can peak at intermediate reconstructions.}
    \label{fig:v2a_bpi_diagnostics}
    \end{minipage}
\end{adjustwidth}

\begin{adjustwidth}{-\extralength}{0cm}
    \begin{minipage}{\fulllength}
    \centering
    \includegraphics[width=\linewidth]{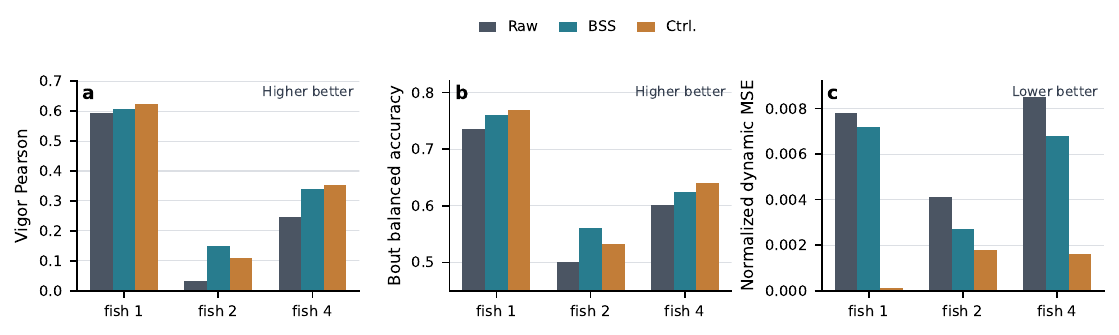}
    \captionof{figure}{Three-recording leakage audited v2a-RSN summary.}
    \label{fig:v2a_strict_three_fish_audit_summary}
    \end{minipage}
\end{adjustwidth}

\subsection{BSS cleaning makes v2a-RSN connectivity matrices denser}
\label{sec:v2a_csl_results}

The final analysis tested whether preprocessing effects reach downstream CSL. For \texttt{fish-1}, \texttt{fish-2} and \texttt{fish-4}, raw traces and \texttt{cluster\_keep\_top\_04} BSS reconstructions were passed through c-GC and c-GC* using \(\tau=2\) and \(n_{\mathrm{pasts}}=4\). The lag \(\tau=2\) was an assumed value corresponding to approximately one-third of the acquisition frequency, and \(n_{\mathrm{pasts}}=4\) was chosen as it is a sufficient conditioning depth for \(\tau=2\); see Adedayo et al. \citep{adedayo2026markovianityDepth}. The \texttt{fish-3} recording was omitted from this CSL analysis because it contains 412 variables. Empirical v2a-RSN PCMCI+ and JPCMCI+ variants were omitted because of runtime and computational cost.

No ground truth v2a-RSN graph is available, so recovery metrics cannot be computed. The sensitivity readout is graph density. Raw traces produced sparse graphs (4.5--6.7\% off diagonal directed edges), whereas BSS cleaned traces were much denser (49.3--87.9\%; Table~\ref{tab:v2a_csl_density}). c-GC and c-GC* gave nearly identical densities, so preprocessing dominated estimator choice.

\section{Discussion}

BSS denoising was not neutral for these calcium traces. In the synthetic benchmark, component-removing BSS variants usually erased recoverable graph structure rather than improving it. In empirical recordings, BSS sometimes improved behavior decoding, but the gains depended on the recording, method, retained clusters and matched controls. It also made downstream c-GC family connectivity matrices far denser than those inferred from raw traces. A preprocessing step that helps one readout can therefore change the graph supplied to causal analysis.

The main practical implication is that denoising should be reported as an intervention on the measured process. Visual cleanup, component plausibility and decoder improvement are insufficient when the downstream target is effective connectivity or effectome analysis \cite{friston2011functional,gilson2016estimation,jelsma2025structural,grosseWentrup2024ncmcm}. For this reason the framework evaluates trace preservation, behavior readouts, time respecting nulls, causal state diagnostics, artefact probes and graph sensitivity together \cite{pospelov2026intense,said2023neurograph}.

BSS should be trusted only when several checks agree. A cleaned trace is more credible when it preserves trace structure, improves behavior readouts beyond matched PCA and low-pass controls, does not reduce causal-state diagnostics to a smoothness gain, passes residual artefact probes and does not induce large graph-density shifts under the intended CSL estimator. If these checks disagree, the cleaned trace is best reported as a sensitivity variant rather than as the default causal input.

The evidence has clear limits. The synthetic benchmark fixes one family of lagged graphs, artefacts and component-selection rules; it demonstrates sensitivity, not universal failure of BSS. Nonfinite values are imputed with whole recording medians before folds, and behavior targets are binned and smoothed before fold construction. The audits therefore cover learned operations, not every input operation. The BSS grid does not vary PCA rank, SOBI lag sets or ICA/Infomax restarts. Prediction level uncertainty is reported, but BPI, causal state, trace preservation and stability summaries lack confidence intervals; best variant summaries remain post hoc upper envelopes. The v2a-RSN connectivity result uses c-GC variants for fish 1, 2 and 4 at fixed \(\tau=2\) and \(n_{\mathrm{pasts}}=4\). The 412-variable fish 3 recording and empirical v2a-RSN PCMCI+/JPCMCI+ variants are omitted because of runtime and computational cost. A fully input-local audit with causal behavior smoothing, lag-grid sensitivity and broader BSS hyperparameter sweeps would be stronger than the current locked analysis.

Within this scope, the conclusion is conservative. BSS may remain useful for artefact exploration and candidate trace reconstruction, but cleaned traces should not be treated as automatically better inputs for behavior linked causal analysis. The safer workflow is to test whether each cleaned representation preserves the evidence required by the intended downstream claim.

%%%%%%%%%%%%%%%%%%%%%%%%%%%%%%%%%%%%%%%%%%
\vspace{6pt}

\supplementary{The supporting information contains motorneuron motivation plots, CSL matrix plots, synthetic trace-preservation diagnostics, consolidated settings, strict-grid summaries, cluster progressions, v2a-RSN CSL matrix plots and detailed method definitions.}

\authorcontributions{Conceptualization, S.A.A.; methodology, S.A.A.; software, S.A.A.; validation, S.A.A.; formal analysis, S.A.A.; investigation, S.A.A.; data curation, S.A.A.; writing---original draft preparation, S.A.A.; writing---review and editing, S.A.A.; visualization, S.A.A.; project administration, S.A.A.; funding acquisition, S.A.A. The author has read and agreed to the published version of the manuscript.}

\funding{This work was funded by the Zebrafish Neuroscience Interdisciplinary Training Hub (ZENITH) program under Marie Curie Actions Agreement grant \#813457.}

\dataavailability{The manuscript analyzes previously extracted larval zebrafish calcium trace recordings and aligned behavior targets supplied for this project. Code and experiment notebooks are available at \url{https://github.com/adedayoas91/ica-denoising}. The repository contains the BSS workflows, leakage-audited behavior and state diagnostics, motorneuron CSL notebooks and v2a-RSN connectivity notebooks used to generate the reported tables and figures.}

\conflictsofinterest{The author declares no conflicts of interest.}

\reftitle{References}
\bibliographystyle{Definitions/mdpi}
\bibliography{references}

\begin{thebibliography}{999}

\bibitem[Ahrens et~al.(2012)Ahrens, Li, Orger, Robson, Schier, Engert, and
  Portugues]{Ahrens2012MotorAdaptation}
Ahrens, M.B.; Li, J.M.; Orger, M.B.; Robson, D.N.; Schier, A.F.; Engert, F.;
  Portugues, R.
\newblock Brain-wide neuronal dynamics during motor adaptation in zebrafish.
\newblock {\em Nature} {\bf 2012}, {\em 485},~471--477.
\newblock {\url{https://doi.org/10.1038/nature11057}}.

\bibitem[Ahrens et~al.(2013)Ahrens, Orger, Robson, Li, and
  Keller]{Ahrens2013WholeBrain}
Ahrens, M.B.; Orger, M.B.; Robson, D.N.; Li, J.M.; Keller, P.J.
\newblock Whole-brain functional imaging at cellular resolution using
  light-sheet microscopy.
\newblock {\em Nature Methods} {\bf 2013}, {\em 10},~413--420.
\newblock {\url{https://doi.org/10.1038/nmeth.2434}}.

\bibitem[Portugues et~al.(2014)Portugues, Feierstein, Engert, and
  Orger]{Portugues2014WholeBrain}
Portugues, R.; Feierstein, C.E.; Engert, F.; Orger, M.B.
\newblock Whole-Brain Activity Maps Reveal Stereotyped, Distributed Networks
  for Visuomotor Behavior.
\newblock {\em Neuron} {\bf 2014}, {\em 81},~1328--1343.
\newblock {\url{https://doi.org/10.1016/j.neuron.2014.01.019}}.

\bibitem[Kim et~al.(2017)Kim, Kim, Marques, Grama, Hildebrand, Gu, Li, and
  Robson]{kim2017pan}
Kim, D.H.; Kim, J.; Marques, J.C.; Grama, A.; Hildebrand, D.G.; Gu, W.; Li,
  J.M.; Robson, D.N.
\newblock Pan-neuronal calcium imaging with cellular resolution in freely
  swimming zebrafish.
\newblock {\em Nature methods} {\bf 2017}, {\em 14},~1107--1114.
\newblock {\url{https://doi.org/10.1038/nmeth.4429}}.

\bibitem[Marques et~al.(2018)Marques, Lackner, F{\'e}lix, and
  Orger]{marques2018structure}
Marques, J.C.; Lackner, S.; F{\'e}lix, R.; Orger, M.B.
\newblock Structure of the zebrafish locomotor repertoire revealed with
  unsupervised behavioral clustering.
\newblock {\em Current Biology} {\bf 2018}, {\em 28},~181--195.

\bibitem[Friston(2011)]{friston2011functional}
Friston, K.J.
\newblock Functional and effective connectivity: a review.
\newblock {\em Brain Connectivity} {\bf 2011}, {\em 1},~13--36.
\newblock {\url{https://doi.org/10.1089/brain.2011.0008}}.

\bibitem[Gilson et~al.(2016)Gilson, Moreno-Bote, Ponce-Alvarez, Ritter, and
  Deco]{gilson2016estimation}
Gilson, M.; Moreno-Bote, R.; Ponce-Alvarez, A.; Ritter, P.; Deco, G.
\newblock Estimation of directed effective connectivity from fMRI functional
  connectivity hints at asymmetries of cortical connectome.
\newblock {\em PLOS Computational Biology} {\bf 2016}, {\em 12},~e1004762.
\newblock {\url{https://doi.org/10.1371/journal.pcbi.1004762}}.

\bibitem[Carroll et~al.(2006)Carroll, Ruppert, Stefanski, and
  Crainiceanu]{carroll2006measurement}
Carroll, R.J.; Ruppert, D.; Stefanski, L.A.; Crainiceanu, C.M.
\newblock {\em Measurement Error in Nonlinear Models: A Modern Perspective}, 2
  ed.; Chapman and Hall/CRC,  2006.
\newblock {\url{https://doi.org/10.1201/9781420010138}}.

\bibitem[Grienberger and Konnerth(2012)]{GrienbergerKonnerth2012Calcium}
Grienberger, C.; Konnerth, A.
\newblock Imaging Calcium in Neurons.
\newblock {\em Neuron} {\bf 2012}, {\em 73},~862--885.
\newblock {\url{https://doi.org/10.1016/j.neuron.2012.02.011}}.

\bibitem[Chen et~al.(2013)Chen, Wardill, Sun, Pulver, Renninger, Baohan,
  Schreiter, Kerr, Orger, Jayaraman, Looger, Svoboda, and Kim]{Chen2013GCaMP}
Chen, T.W.; Wardill, T.J.; Sun, Y.; Pulver, S.R.; Renninger, S.L.; Baohan, A.;
  Schreiter, E.R.; Kerr, R.A.; Orger, M.B.; Jayaraman, V.;  et~al.
\newblock Ultrasensitive fluorescent proteins for imaging neuronal activity.
\newblock {\em Nature} {\bf 2013}, {\em 499},~295--300.
\newblock {\url{https://doi.org/10.1038/nature12354}}.

\bibitem[Pnevmatikakis and Giovannucci(2017)]{pnevmatikakis2017normcorre}
Pnevmatikakis, E.A.; Giovannucci, A.
\newblock NoRMCorre: An online algorithm for piecewise rigid motion correction
  of calcium imaging data.
\newblock {\em Journal of Neuroscience Methods} {\bf 2017}, {\em 291},~83--94.
\newblock {\url{https://doi.org/10.1016/j.jneumeth.2017.07.031}}.

\bibitem[Pnevmatikakis et~al.(2016)Pnevmatikakis, Soudry, Gao, Machado, Merel,
  Pfau, Reardon, Mu, Lacefield, Yang, Ahrens, Bruno, Jessell, Peterka, Yuste,
  and Paninski]{Pnevmatikakis2016Calcium}
Pnevmatikakis, E.A.; Soudry, D.; Gao, Y.; Machado, T.A.; Merel, J.; Pfau, D.;
  Reardon, T.; Mu, Y.; Lacefield, C.; Yang, W.;  et~al.
\newblock Simultaneous Denoising, Deconvolution, and Demixing of Calcium
  Imaging Data.
\newblock {\em Neuron} {\bf 2016}, {\em 89},~285--299.
\newblock {\url{https://doi.org/10.1016/j.neuron.2015.11.037}}.

\bibitem[Giovannucci et~al.(2019)Giovannucci, Friedrich, Gunn, Kalfon, Brown,
  Koay, Taxidis, Najafi, Gauthier, Zhou, et~al.]{giovannucci2019caiman}
Giovannucci, A.; Friedrich, J.; Gunn, P.; Kalfon, J.; Brown, B.L.; Koay, S.A.;
  Taxidis, J.; Najafi, F.; Gauthier, J.L.; Zhou, P.;  et~al.
\newblock CaImAn an open source tool for scalable calcium imaging data
  analysis.
\newblock {\em elife} {\bf 2019}, {\em 8},~e38173.
\newblock {\url{https://doi.org/10.7554/eLife.38173}}.

\bibitem[Pnevmatikakis(2019)]{pnevmatikakis2019analysis}
Pnevmatikakis, E.A.
\newblock Analysis pipelines for calcium imaging data.
\newblock {\em Current opinion in neurobiology} {\bf 2019}, {\em 55},~15--21.
\newblock {\url{https://doi.org/10.1016/j.conb.2018.11.004}}.

\bibitem[Kim and Schnitzer(2022)]{kim2022fluorescence}
Kim, T.H.; Schnitzer, M.J.
\newblock Fluorescence imaging of large-scale neural ensemble dynamics.
\newblock {\em Cell} {\bf 2022}, {\em 185},~9--41.

\bibitem[Vanwalleghem et~al.(2021)Vanwalleghem, Constantin, and
  Scott]{vanwalleghem2021calcium}
Vanwalleghem, G.; Constantin, L.; Scott, E.K.
\newblock Calcium imaging and the curse of negativity.
\newblock {\em Frontiers in neural circuits} {\bf 2021}, {\em 14},~607391.

\bibitem[Spirtes et~al.(2000)Spirtes, Glymour, Scheines, and
  Heckerman]{spirtes2000causation}
Spirtes, P.; Glymour, C.N.; Scheines, R.; Heckerman, D.
\newblock {\em Causation, prediction, and search}; MIT press,  2000.

\bibitem[Pearl(2009)]{pearl2009causality}
Pearl, J.
\newblock {\em Causality}; Cambridge university press,  2009.

\bibitem[Peters et~al.(2017)Peters, Janzing, and
  Sch{\"o}lkopf]{peters2017elements}
Peters, J.; Janzing, D.; Sch{\"o}lkopf, B.
\newblock {\em Elements of causal inference: foundations and learning
  algorithms}; The MIT Press,  2017.

\bibitem[Runge et~al.(2023)Runge, Gerhardus, Varando, Eyring, and
  Camps-Valls]{runge2023causal}
Runge, J.; Gerhardus, A.; Varando, G.; Eyring, V.; Camps-Valls, G.
\newblock Causal inference for time series.
\newblock {\em Nature Reviews Earth \& Environment} {\bf 2023}, {\em
  4},~487--505.

\bibitem[Florin et~al.(2010)Florin, Gross, Pfeifer, Fink, and
  Timmermann]{florin2010filtering}
Florin, E.; Gross, J.; Pfeifer, J.; Fink, G.R.; Timmermann, L.
\newblock The effect of filtering on Granger causality based multivariate
  causality measures.
\newblock {\em NeuroImage} {\bf 2010}, {\em 50},~577--588.
\newblock {\url{https://doi.org/10.1016/j.neuroimage.2009.12.050}}.

\bibitem[Barnett and Seth(2011)]{barnett2011filtering}
Barnett, L.; Seth, A.K.
\newblock Behaviour of Granger causality under filtering: Theoretical
  invariance and practical application.
\newblock {\em Journal of Neuroscience Methods} {\bf 2011}, {\em
  201},~404--419.
\newblock {\url{https://doi.org/10.1016/j.jneumeth.2011.08.010}}.

\bibitem[Seth et~al.(2015)Seth, Barrett, and Barnett]{Seth3293}
Seth, A.K.; Barrett, A.B.; Barnett, L.
\newblock Granger Causality Analysis in Neuroscience and Neuroimaging.
\newblock {\em Journal of Neuroscience} {\bf 2015}, {\em 35},~3293--3297,
  \href{http://arxiv.org/abs/https://www.jneurosci.org/content/35/8/3293.full.pdf}{{\normalfont
  [https://www.jneurosci.org/content/35/8/3293.full.pdf]}}.
\newblock {\url{https://doi.org/10.1523/JNEUROSCI.4399-14.2015}}.

\bibitem[Seth et~al.(2013)Seth, Chorley, and Barnett]{seth2013fmri}
Seth, A.K.; Chorley, P.; Barnett, L.C.
\newblock Granger causality analysis of fMRI BOLD signals is invariant to
  hemodynamic convolution but not downsampling.
\newblock {\em NeuroImage} {\bf 2013}, {\em 65},~540--555.
\newblock {\url{https://doi.org/10.1016/j.neuroimage.2012.09.049}}.

\bibitem[Weber et~al.(2017)Weber, Florin, von Papen, and
  Timmermann]{weber2017transfer}
Weber, I.; Florin, E.; von Papen, M.; Timmermann, L.
\newblock The influence of filtering and downsampling on the estimation of
  transfer entropy.
\newblock {\em PLOS ONE} {\bf 2017}, {\em 12},~e0188210.
\newblock {\url{https://doi.org/10.1371/journal.pone.0188210}}.

\bibitem[Luo et~al.(2013)Luo, Lu, Cheng, Valdes-Sosa, Wen, Ding, and
  Feng]{luo2013spatiotemporal}
Luo, Q.; Lu, W.; Cheng, W.; Valdes-Sosa, P.A.; Wen, X.; Ding, M.; Feng, J.
\newblock Spatio-temporal Granger causality: A new framework.
\newblock {\em NeuroImage} {\bf 2013}, {\em 79},~241--263.
\newblock {\url{https://doi.org/10.1016/j.neuroimage.2013.04.091}}.

\bibitem[Zhou et~al.(2014)Zhou, Zhang, Xiao, and Cai]{zhou2014sampling}
Zhou, D.; Zhang, Y.; Xiao, Y.; Cai, D.
\newblock Analysis of sampling artifacts on the Granger causality analysis for
  topology extraction of neuronal dynamics.
\newblock {\em Frontiers in Computational Neuroscience} {\bf 2014}, {\em
  8},~75.
\newblock {\url{https://doi.org/10.3389/fncom.2014.00075}}.

\bibitem[Barnett and Seth(2017)]{barnett2017subsampling}
Barnett, L.; Seth, A.K.
\newblock Detectability of Granger causality for subsampled continuous-time
  neurophysiological processes.
\newblock {\em Journal of Neuroscience Methods} {\bf 2017}, {\em 275},~93--121.
\newblock {\url{https://doi.org/10.1016/j.jneumeth.2016.10.016}}.

\bibitem[Gong et~al.(2017)Gong, Zhang, Sch{\"o}lkopf, Glymour, and
  Tao]{gong2017temporal}
Gong, M.; Zhang, K.; Sch{\"o}lkopf, B.; Glymour, C.; Tao, D.
\newblock Causal discovery from temporally aggregated time series.
\newblock In Proceedings of the Proceedings of the 33rd Conference on
  Uncertainty in Artificial Intelligence,  2017, p. 269.

\bibitem[Hyttinen et~al.(2017)Hyttinen, Plis, J{\"a}rvisalo, Eberhardt, and
  Danks]{hyttinen2017subsampled}
Hyttinen, A.; Plis, S.; J{\"a}rvisalo, M.; Eberhardt, F.; Danks, D.
\newblock A constraint optimization approach to causal discovery from
  subsampled time series data.
\newblock {\em International Journal of Approximate Reasoning} {\bf 2017}, {\em
  90},~208--225.
\newblock {\url{https://doi.org/10.1016/j.ijar.2017.07.009}}.

\bibitem[Plis et~al.(2015)Plis, Danks, Freeman, and
  Calhoun]{plis2015rateagnostic}
Plis, S.; Danks, D.; Freeman, C.; Calhoun, V.
\newblock Rate-Agnostic (Causal) Structure Learning.
\newblock In Proceedings of the Advances in Neural Information Processing
  Systems,  2015, Vol.~28.

\bibitem[Hyvarinen(1999)]{Hyvarinen1999FastICA}
Hyvarinen, A.
\newblock Fast and Robust Fixed-Point Algorithms for Independent Component
  Analysis.
\newblock {\em IEEE Transactions on Neural Networks} {\bf 1999}, {\em
  10},~626--634.
\newblock {\url{https://doi.org/10.1109/72.761722}}.

\bibitem[Bell and Sejnowski(1995)]{BellSejnowski1995Infomax}
Bell, A.J.; Sejnowski, T.J.
\newblock An Information-Maximization Approach to Blind Separation and Blind
  Deconvolution.
\newblock {\em Neural Computation} {\bf 1995}, {\em 7},~1129--1159.
\newblock {\url{https://doi.org/10.1162/neco.1995.7.6.1129}}.

\bibitem[Belouchrani et~al.(1997)Belouchrani, Abed-Meraim, Cardoso, and
  Moulines]{belouchrani1997sobi}
Belouchrani, A.; Abed-Meraim, K.; Cardoso, J.F.; Moulines, E.
\newblock A blind source separation technique using second-order statistics.
\newblock {\em IEEE Transactions on Signal Processing} {\bf 1997}, {\em
  45},~434--444.
\newblock {\url{https://doi.org/10.1109/78.554307}}.

\bibitem[Cardoso and Souloumiac(1993)]{CardosoSouloumiac1993JADE}
Cardoso, J.F.; Souloumiac, A.
\newblock Blind Beamforming for Non Gaussian Signals.
\newblock {\em IEE Proceedings F: Radar and Signal Processing} {\bf 1993}, {\em
  140},~362--370.
\newblock {\url{https://doi.org/10.1049/ip-f-2.1993.0054}}.

\bibitem[Hyvarinen and Oja(2000)]{HyvarinenOja2000ICA}
Hyvarinen, A.; Oja, E.
\newblock Independent Component Analysis: Algorithms and Applications.
\newblock {\em Neural Networks} {\bf 2000}, {\em 13},~411--430.
\newblock {\url{https://doi.org/10.1016/S0893-6080(00)00026-5}}.

\bibitem[Brown et~al.(2001)Brown, Yamada, and
  Sejnowski]{BrownYamadaSejnowski2001NeuralICA}
Brown, G.D.; Yamada, S.; Sejnowski, T.J.
\newblock Independent component analysis at the neural cocktail party.
\newblock {\em Trends in Neurosciences} {\bf 2001}, {\em 24},~54--63.
\newblock {\url{https://doi.org/10.1016/S0166-2236(00)01683-0}}.

\bibitem[Vigario and Oja(2008)]{VigarioOja2008Neuroinformatics}
Vigario, R.; Oja, E.
\newblock BSS and ICA in Neuroinformatics: From Current Practices to Open
  Challenges.
\newblock {\em IEEE Reviews in Biomedical Engineering} {\bf 2008}, {\em
  1},~50--61.
\newblock {\url{https://doi.org/10.1109/RBME.2008.2008244}}.

\bibitem[Mukamel et~al.(2009)Mukamel, Nimmerjahn, and
  Schnitzer]{Mukamel2009CaImaging}
Mukamel, E.A.; Nimmerjahn, A.; Schnitzer, M.J.
\newblock Automated Analysis of Cellular Signals from Large-Scale Calcium
  Imaging Data.
\newblock {\em Neuron} {\bf 2009}, {\em 63},~747--760.
\newblock {\url{https://doi.org/10.1016/j.neuron.2009.08.009}}.

\bibitem[Chen et~al.(2023)Chen, Ginoux, Carbo-Tano, Mora, Walczak, and
  Wyart]{chen2023granger}
Chen, X.; Ginoux, F.; Carbo-Tano, M.; Mora, T.; Walczak, A.M.; Wyart, C.
\newblock Granger causality analysis for calcium transients in neuronal
  networks, challenges and improvements.
\newblock {\em eLife} {\bf 2023}, {\em 12},~e81279.
\newblock {\url{https://doi.org/10.7554/eLife.81279}}.

\bibitem[Fallani et~al.(2014)Fallani, Corazzol, Sternberg, Wyart, and
  Chavez]{fallani2014hierarchy}
Fallani, F.D.V.; Corazzol, M.; Sternberg, J.R.; Wyart, C.; Chavez, M.
\newblock Hierarchy of neural organization in the embryonic spinal cord:
  Granger-causality graph analysis of in vivo calcium imaging data.
\newblock {\em IEEE Transactions on Neural Systems and Rehabilitation
  Engineering} {\bf 2014}, {\em 23},~333--341.
\newblock {\url{https://doi.org/10.1109/TNSRE.2014.2341632}}.

\bibitem[Carbo-Tano et~al.(2022)Carbo-Tano, Lapoix, Jia, Auclair, Dubuc, and
  Wyart]{carbo2022functional}
Carbo-Tano, M.; Lapoix, M.; Jia, X.; Auclair, F.; Dubuc, R.; Wyart, C.
\newblock Functional coupling of the mesencephalic locomotor region and v2a
  reticulospinal neurons driving forward locomotion.
\newblock {\em bioRxiv} {\bf 2022}, pp. 2022--04.
\newblock {\url{https://doi.org/10.1101/2022.04.01.486703}}.

\bibitem[Shen et~al.(2021)Shen, Lur, Xu, and Yu]{shen2021deconvolve}
Shen, T.; Lur, G.; Xu, X.; Yu, Z.
\newblock To Deconvolve, or Not to Deconvolve: Inferences of Neuronal
  Activities using Calcium Imaging Data.
\newblock {\em arXiv preprint arXiv:2103.02163} {\bf 2021},
  \href{http://arxiv.org/abs/2103.02163}{{\normalfont
  [arXiv:q-bio.NC/2103.02163]}}.

\bibitem[Welch(1967)]{Welch1967PSD}
Welch, P.D.
\newblock The Use of Fast Fourier Transform for the Estimation of Power
  Spectra: A Method Based on Time Averaging Over Short, Modified Periodograms.
\newblock {\em IEEE Transactions on Audio and Electroacoustics} {\bf 1967},
  {\em 15},~70--73.
\newblock {\url{https://doi.org/10.1109/TAU.1967.1161901}}.

\bibitem[Lloyd(1982)]{Lloyd1982KMeans}
Lloyd, S.P.
\newblock Least Squares Quantization in PCM.
\newblock {\em IEEE Transactions on Information Theory} {\bf 1982}, {\em
  28},~129--137.
\newblock {\url{https://doi.org/10.1109/TIT.1982.1056489}}.

\bibitem[Kumar et~al.(2023)Kumar, Gilra, Gonzalez-Soto, Meunier, and
  Grosse-Wentrup]{kumar2023bundle}
Kumar, A.; Gilra, A.; Gonzalez-Soto, M.; Meunier, A.; Grosse-Wentrup, M.
\newblock BunDLe-Net: Neuronal Manifold Learning Meets Behaviour.
\newblock {\em bioRxiv} {\bf 2023}.
\newblock {\url{https://doi.org/10.1101/2023.08.08.551978}}.

\bibitem[Grosse-Wentrup et~al.(2024)Grosse-Wentrup, Kumar, Meunier, and
  Zimmer]{grosseWentrup2024ncmcm}
Grosse-Wentrup, M.; Kumar, A.; Meunier, A.; Zimmer, M.
\newblock Neuro-cognitive multilevel causal modeling: A framework that bridges
  the explanatory gap between neuronal activity and cognition.
\newblock {\em PLOS Computational Biology} {\bf 2024}, {\em 20},~e1012674.
\newblock {\url{https://doi.org/10.1371/journal.pcbi.1012674}}.

\bibitem[Tipping and Bishop(1999)]{TippingBishop1999PCA}
Tipping, M.E.; Bishop, C.M.
\newblock Probabilistic Principal Component Analysis.
\newblock {\em Journal of the Royal Statistical Society Series B: Statistical
  Methodology} {\bf 1999}, {\em 61},~611--622.
\newblock {\url{https://doi.org/10.1111/1467-9868.00196}}.

\bibitem[Varoquaux et~al.(2017)Varoquaux, Raamana, Engemann, Hoyos-Idrobo,
  Schwartz, and Thirion]{Varoquaux2017DecoderCV}
Varoquaux, G.; Raamana, P.R.; Engemann, D.A.; Hoyos-Idrobo, A.; Schwartz, Y.;
  Thirion, B.
\newblock Assessing and tuning brain decoders: Cross-validation, caveats, and
  guidelines.
\newblock {\em NeuroImage} {\bf 2017}, {\em 145},~166--179.
\newblock {\url{https://doi.org/10.1016/j.neuroimage.2016.10.038}}.

\bibitem[Vabalas et~al.(2019)Vabalas, Gowen, Poliakoff, and
  Casson]{Vabalas2019LimitedSample}
Vabalas, A.; Gowen, E.; Poliakoff, E.; Casson, A.J.
\newblock Machine learning algorithm validation with a limited sample size.
\newblock {\em PLOS ONE} {\bf 2019}, {\em 14},~e0224365.
\newblock {\url{https://doi.org/10.1371/journal.pone.0224365}}.

\bibitem[Cawley and Talbot(2010)]{CawleyTalbot2010SelectionBias}
Cawley, G.C.; Talbot, N.L.C.
\newblock On Over-fitting in Model Selection and Subsequent Selection Bias in
  Performance Evaluation.
\newblock {\em Journal of Machine Learning Research} {\bf 2010}, {\em
  11},~2079--2107.

\bibitem[Adedayo(2025)]{adedayo2025cgc}
Adedayo, S.A.
\newblock Re-examining Granger Causality with Causal Bayesian Networks and
  Reichenbachs Principles,  2025,
  \href{http://arxiv.org/abs/2501.02672}{{\normalfont
  [arXiv:stat.ML/2501.02672]}}.
\newblock arXiv:2501.02672, revised May 2026.

\bibitem[Runge et~al.(2019)Runge, Nowack, Kretschmer, Flaxman, and
  Sejdinovic]{runge2019detecting}
Runge, J.; Nowack, P.; Kretschmer, M.; Flaxman, S.; Sejdinovic, D.
\newblock Detecting and quantifying causal associations in large nonlinear time
  series datasets.
\newblock {\em Science advances} {\bf 2019}, {\em 5},~eaau4996.

\bibitem[G{\"u}nther et~al.(2023)G{\"u}nther, Ninad, and
  Runge]{gunther2023jpcmci}
G{\"u}nther, W.; Ninad, U.; Runge, J.
\newblock Causal Discovery for Time Series from Multiple Datasets with Latent
  Contexts.
\newblock In Proceedings of the Proceedings of the Thirty-Ninth Conference on
  Uncertainty in Artificial Intelligence; Evans, R.J.; Shpitser, I., Eds. PMLR,
   2023, Vol. 216, {\em Proceedings of Machine Learning Research}, pp.
  766--776.

\bibitem[Adedayo(2026)]{adedayo2026markovianityDepth}
Adedayo, S.A.
\newblock Conditioning-Depth Diagnostics for Hidden Memory in Temporal Causal
  Discovery.
\newblock {\em arXiv preprint arXiv:2606.01214} {\bf 2026},
  \href{http://arxiv.org/abs/2606.01214}{{\normalfont
  [arXiv:stat.AP/2606.01214]}}.

\bibitem[Jelsma et~al.(2025)Jelsma, Zijlmans, Heijink, Hoefnagels, Raemaekers,
  Otte, van Klink, and van Blooijs]{jelsma2025structural}
Jelsma, S.B.; Zijlmans, M.; Heijink, I.B.; Hoefnagels, F.W.A.; Raemaekers, M.;
  Otte, W.M.; van Klink, N.E.C.; van Blooijs, D.
\newblock Structural and effective brain connectivity in focal epilepsy.
\newblock {\em NeuroImage: Reports} {\bf 2025}, {\em 5},~100274.
\newblock {\url{https://doi.org/10.1016/j.ynirp.2025.100274}}.

\bibitem[Pospelov et~al.(2026)Pospelov, Plusnin, Rogozhnikova, Ivanova,
  Sotskov, Toropova, Ivashkina, Avetisov, and Anokhin]{pospelov2026intense}
Pospelov, N.; Plusnin, V.; Rogozhnikova, O.; Ivanova, A.; Sotskov, V.;
  Toropova, K.; Ivashkina, O.; Avetisov, V.; Anokhin, K.
\newblock {INTENSE}: Detecting and disentangling neuronal selectivity in
  calcium imaging data.
\newblock {\em arXiv preprint arXiv:2603.04622} {\bf 2026},
  \href{http://arxiv.org/abs/2603.04622}{{\normalfont
  [arXiv:q-bio.NC/2603.04622]}}.

\bibitem[Said et~al.(2023)Said, Bayrak, Derr, Shabbir, Moyer, Chang, and
  Koutsoukos]{said2023neurograph}
Said, A.; Bayrak, R.G.; Derr, T.; Shabbir, M.; Moyer, D.; Chang, C.;
  Koutsoukos, X.
\newblock {NeuroGraph}: Benchmarks for Graph Machine Learning in Brain
  Connectomics.
\newblock {\em Advances in Neural Information Processing Systems} {\bf 2023},
  {\em 36},~65073--65097,  \href{http://arxiv.org/abs/2306.06202}{{\normalfont
  [arXiv:cs.LG/2306.06202]}}.

\bibitem[Cardoso and Souloumiac(1996)]{CardosoSouloumiac1996Jacobi}
Cardoso, J.F.; Souloumiac, A.
\newblock Jacobi Angles for Simultaneous Diagonalization.
\newblock {\em SIAM Journal on Matrix Analysis and Applications} {\bf 1996},
  {\em 17},~161--164.
\newblock {\url{https://doi.org/10.1137/S0895479893259546}}.

\bibitem[Hoyer(2004)]{Hoyer2004Sparsity}
Hoyer, P.O.
\newblock Non-negative Matrix Factorization with Sparseness Constraints.
\newblock {\em Journal of Machine Learning Research} {\bf 2004}, {\em
  5},~1457--1469.

\end{thebibliography}

\clearpage
\begin{suppwide}
\noindent\section*{Supplementary Materials}
\end{suppwide}
\setcounter{section}{0}
\setcounter{subsection}{0}
\renewcommand{\thesection}{S\arabic{section}}
\renewcommand{\thesubsection}{S\arabic{section}.\arabic{subsection}}
\renewcommand{\thefigure}{S\arabic{figure}}
\renewcommand{\thetable}{S\arabic{table}}
\renewcommand{\theHsection}{S\arabic{section}}
\renewcommand{\theHsubsection}{S\arabic{section}.\arabic{subsection}}
\renewcommand{\theHfigure}{S\arabic{figure}}
\renewcommand{\theHtable}{S\arabic{table}}
\setcounter{figure}{0}
\setcounter{table}{0}
\section{Motorneuron motivation and CSL sensitivity}
\label{sec:supp_motorneuron_trace_reconstructions}

The motorneuron recording is used only as an inspectable motivation and
sensitivity check, not as a main empirical result. It contains a visible
artefact at frame~139 and has a physiological ipsilateral rostral-to-caudal
hypothesis, but no validated edge-level ground truth
\cite{chen2023granger, fallani2014hierarchy}.

\begin{suppwide}
    \centering
    \begin{minipage}[t]{0.57\suppdisplaywidth}
        \vspace{0pt}
        \centering
        \includegraphics[width=\linewidth]{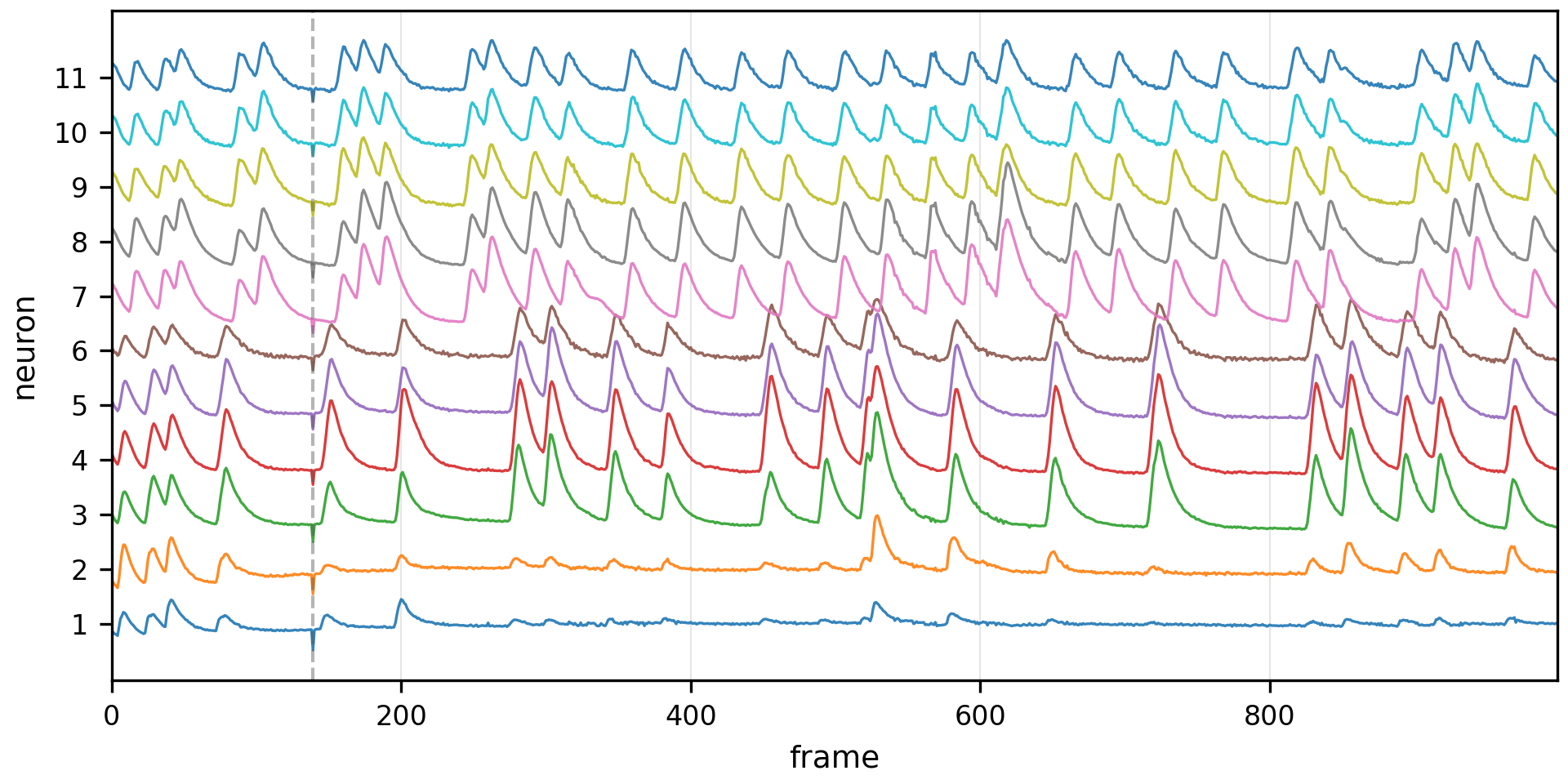}
        \captionof{figure}{Raw motorneuron recording used as motivation. The
        gray vertical line marks frame~139.}
        \label{fig:supp_motorneuron_motivation}
    \end{minipage}
    \hfill
    \begin{minipage}[t]{0.40\suppdisplaywidth}
        \vspace{0pt}
        \centering
        \captionof{table}{Motorneuron CSL sensitivity to BSS preprocessing.
        Entries are off diagonal directed edge counts.}
        \label{tab:supp_motorneuron_csl_edge_counts}
        \scriptsize
        \setlength{\tabcolsep}{2.2pt}
        \resizebox{\linewidth}{!}{%
        \begin{tabular}{lrrrrr}
        \toprule
        Learners & Raw & FastICA & Infomax & SOBI & JADE \\
        \midrule
        c-GC     & 36 & 62 & 74 & 94 & 70 \\
        c-GC*    & 30 & 56 & 61 & 81 & 59 \\
        PCMCI+   & 20 &  8 &  4 &  0 & 15 \\
        JPCMCI+  & 20 &  8 &  4 &  0 & 15 \\
        \bottomrule
        \end{tabular}
        }
    \end{minipage}
\end{suppwide}
\vspace{0.75\baselineskip}

Figure~\ref{fig:supp_motorneuron_cleaned_variants} expands the cleaned trace
reconstructions across all four BSS methods.

\begin{suppwide}
    \begin{minipage}{\suppdisplaywidth}
        \centering
        \includegraphics[width=\linewidth]{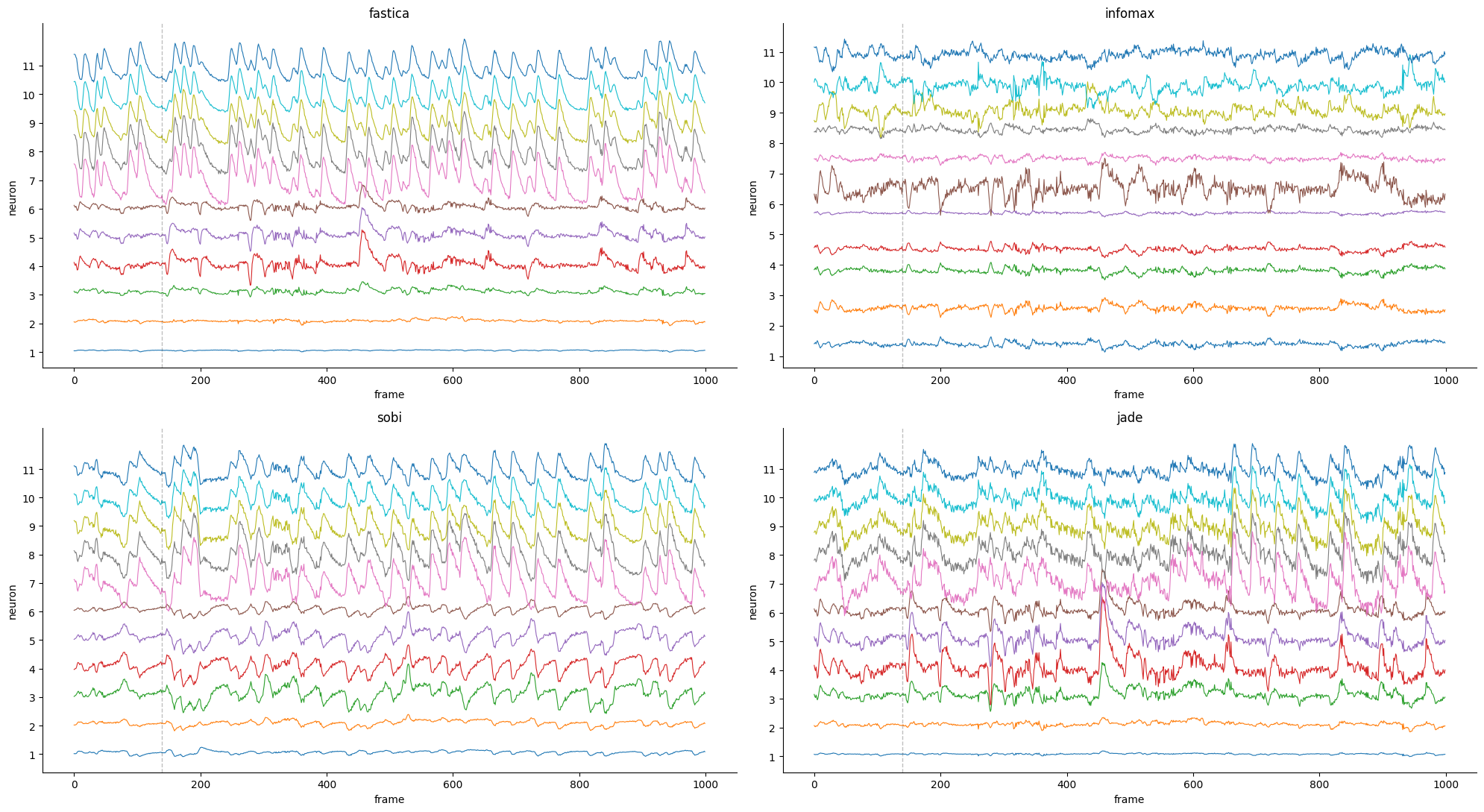}
        \captionof{figure}{BSS cleaned motorneuron traces for FastICA, Infomax, SOBI, and
        JADE. The gray vertical line marks frame~139.}
        \label{fig:supp_motorneuron_cleaned_variants}
    \end{minipage}
\end{suppwide}

\subsection{Motorneuron CSL plots}
\label{sec:supp_motorneuron_csl_artifacts}

Table~\ref{tab:supp_motorneuron_csl_edge_counts} summarizes the same
motorneuron causal structure learning outputs by directed edge count.
Figure~\ref{fig:supp_motorneuron_csl_all} shows the corresponding summaries for
c-GC, c-GC*, PCMCI+, and JPCMCI+. Each grid compares the raw trace with the
FastICA, Infomax, SOBI, and JADE cleaned variants. The top row shows adjacency
matrices, where rows are source neurons and columns are target neurons; the
bottom row redraws the same matrices as directed graphs in the bilateral
motorneuron layout.

\begin{suppwide}
    \centering
    \begin{minipage}{\suppdisplaywidth}
    \centering
    \begin{minipage}[t]{0.49\linewidth}
        \centering
        \includegraphics[width=\linewidth]{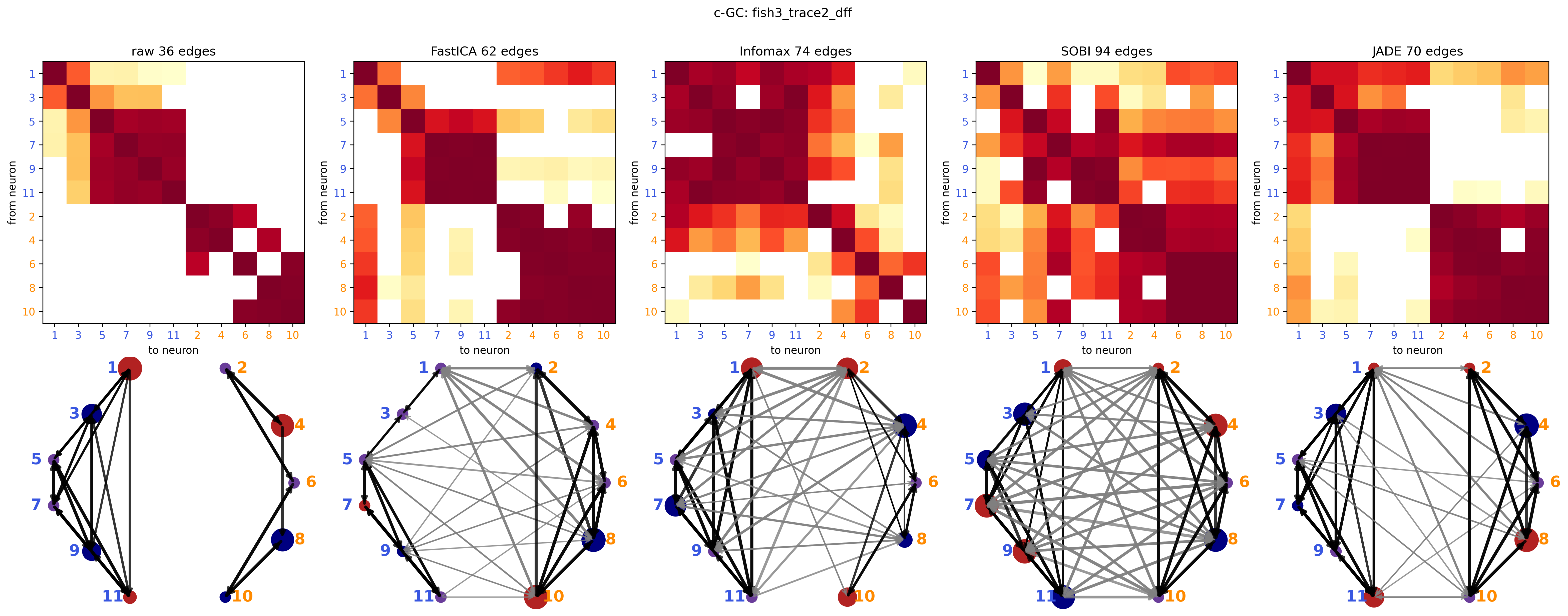}
        {\footnotesize\textbf{a.} c-GC.\par}
    \end{minipage}
    \hfill
    \begin{minipage}[t]{0.49\linewidth}
        \centering
        \includegraphics[width=\linewidth]{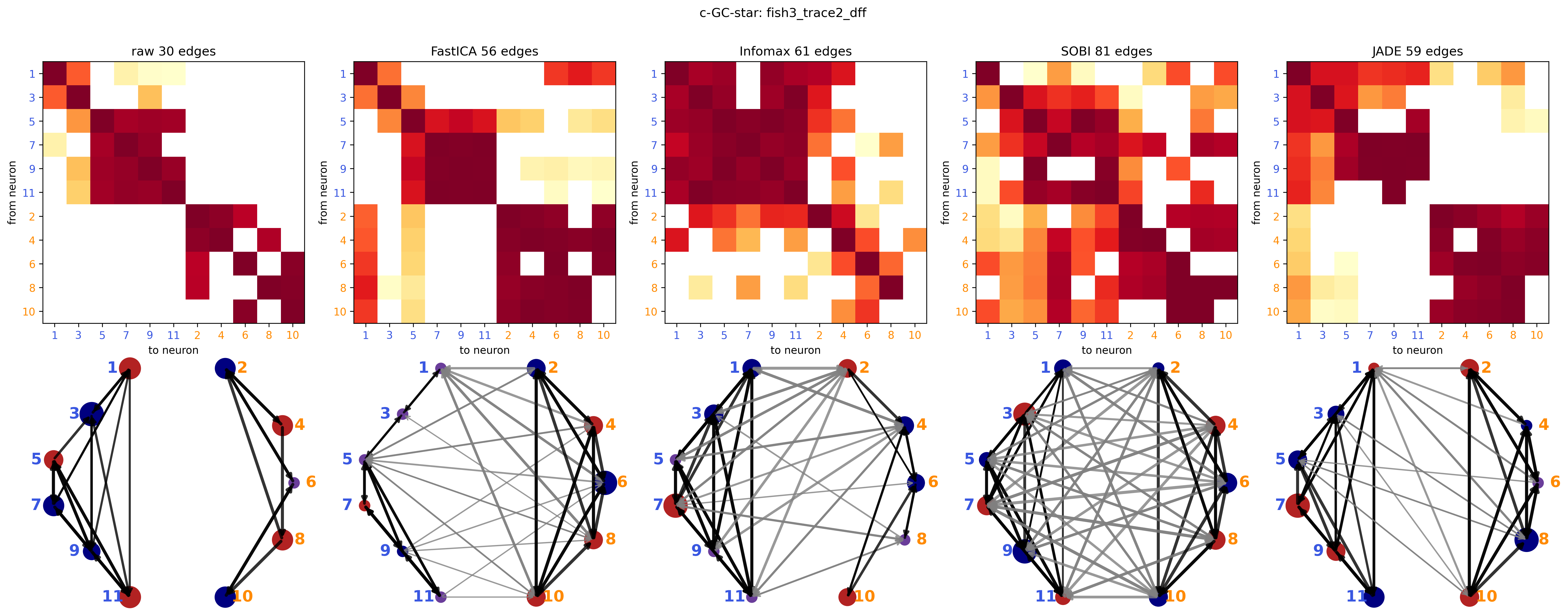}
        {\footnotesize\textbf{b.} c-GC*.\par}
    \end{minipage}

    \vspace{0.4em}
    \begin{minipage}[t]{0.49\linewidth}
        \centering
        \includegraphics[width=\linewidth]{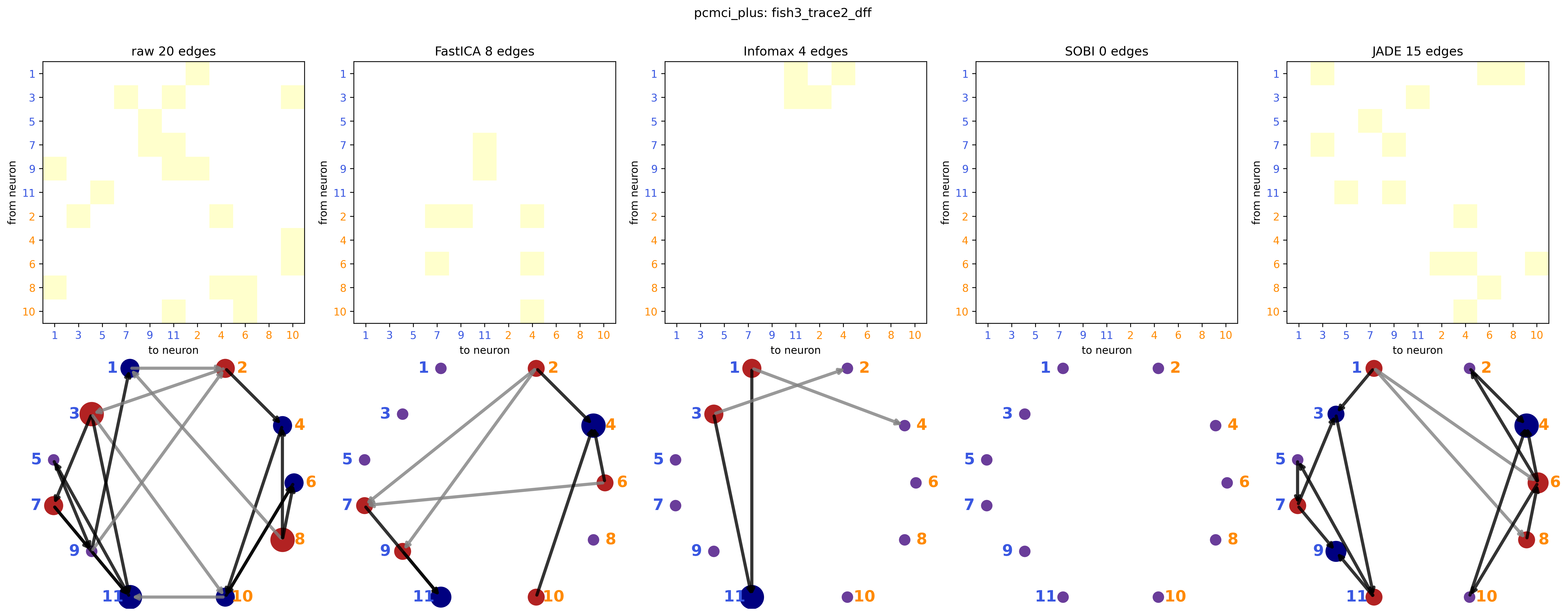}
        {\footnotesize\textbf{c.} PCMCI+.\par}
    \end{minipage}
    \hfill
    \begin{minipage}[t]{0.49\linewidth}
        \centering
        \includegraphics[width=\linewidth]{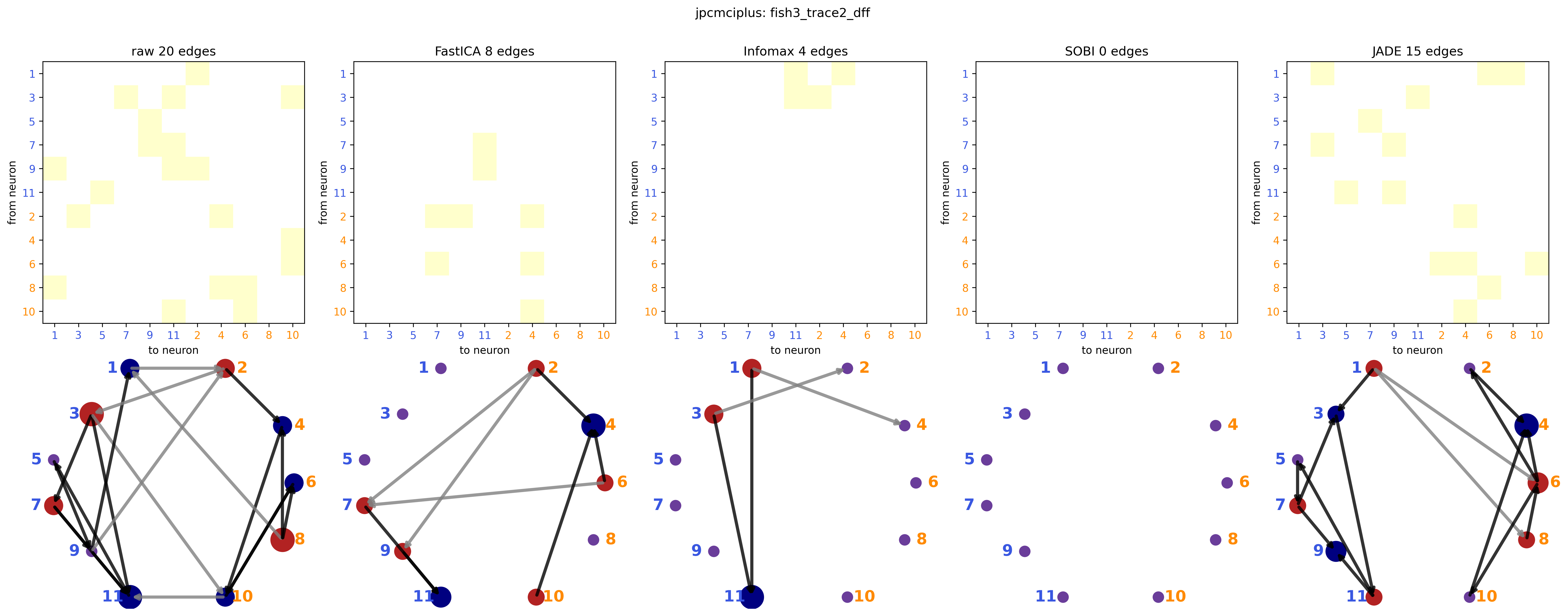}
        {\footnotesize\textbf{d.} JPCMCI+.\par}
    \end{minipage}
    \captionof{figure}{Motorneuron CSL adjacency matrices and directed graphs.}
    \label{fig:supp_motorneuron_csl_all}
    \end{minipage}
\end{suppwide}
\vspace{0.75\baselineskip}

Label color marks side, with blue labels for the
odd numbered neurons and orange labels for the even numbered neurons. Node fill
summarizes the net ipsilateral balance used for the graph view: red nodes have
more outgoing than incoming ipsilateral links, blue nodes have more incoming
than outgoing ipsilateral links, and purple nodes are approximately balanced;
larger nodes indicate a larger imbalance. Black arrows connect ipsilateral
neurons, gray arrows cross between sides, and arrow thickness follows the
corresponding matrix entry.

\subsection{Synthetic trace preservation diagnostic}
\label{sec:supp_simulation_trace_graph}

Trace preservation and graph recovery were not interchangeable in the synthetic
benchmark. Raw traces had a median trace correlation of 0.83 against clean
fluorescence, whereas the component-removing BSS variants had a median trace
correlation of 0.60.

\begin{suppwide}
    \begin{minipage}{\suppdisplaywidth}
        \centering
        \includegraphics[width=0.82\linewidth]{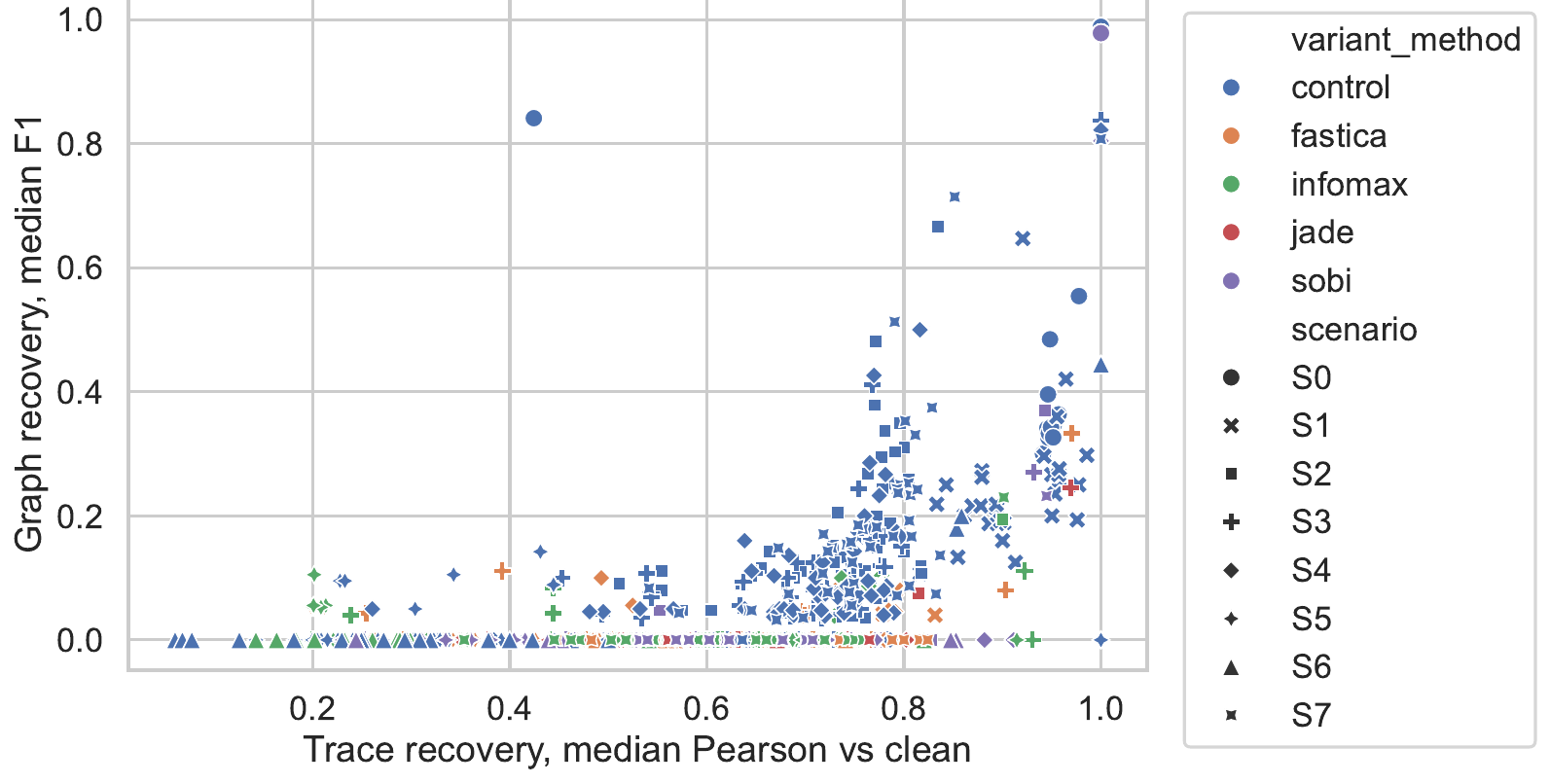}
        \captionof{figure}{Trace preservation and graph recovery in the synthetic benchmark.
        Each point is a trace variant summarized by median trace correlation against
        clean trace and median graph F1.}
        \label{fig:supp_simulation_trace_vs_graph_recovery}
    \end{minipage}
\end{suppwide}

\subsection{Consolidated settings}
\label{sec:supp_execution_settings}

Table~\ref{tab:supp_strict_settings} centralizes the algorithmic choices that
otherwise appear across decomposition, selection, decoding, and stability
analyses. The current configuration is the reported three-recording fold-local
grid used for the v2a-RSN results. It is leakage audited for learned operations
in \texttt{fish-1}, \texttt{fish-2}, and \texttt{fish-4}; \texttt{fish-3} is
omitted because of computational cost. It remains
limited by nonfinite value imputation and behavior smoothing before splitting,
plus convergence warnings. Values recorded in the per run manifests take
precedence over software defaults, so Table~\ref{tab:supp_strict_settings}
should be read as the reporting index for the reported outputs rather than as a
general default configuration.

\begin{suppwide}
\begin{minipage}{\suppdisplaywidth}
\centering
\footnotesize
\renewcommand{\arraystretch}{0.95}
\captionof{table}{Consolidated fold local evaluation settings.}
\label{tab:supp_strict_settings}
\begin{tabular}{p{0.16\linewidth}p{0.45\linewidth}p{0.25\linewidth}}
\hline
\shortstack{Stage} & \shortstack{Current\\configuration} & \shortstack{Output\\notes} \\
\hline
Evaluation boundary &
Five contiguous test blocks; 14-frame exclusion gap; disjoint training
segments preserved for lagged covariance and Welch features. &
Learned operation leakage audits pass in the three reported v2a-RSN recordings. \\
BSS methods and rank &
FastICA, Infomax, SOBI, and approximate JADE; full trace count rank for the
current recordings; tolerance \(10^{-4}\); maximum 500 iterations; random
state 0. &
Some BSS fits in the reported recordings produce nonconvergence warnings. \\
Joint diagonalization &
SOBI lag set \(\{1,2,3,5\}\); JADE capped at 200 cumulant matrices; shared
Jacobi style joint diagonalizer. &
Exercised for SOBI and JADE in the current outputs. \\
Spectral feature clustering &
Seven K-means clusters; Welch \texttt{nperseg}=250 and
\texttt{noverlap}=125; discard bins before index 30; \(\log(1+P)\) features;
cluster rank is peak mean log PSD in 0--0.20~Hz. &
Cluster assignments, component selections, and stability summaries are written
per recording. \\
Selection controls &
Retain 1--7 clusters under low frequency, high frequency, random,
energy matched random, and all component rules; random seeds 0--4. &
Selection rule sensitivity is part of the reported limitation. \\
Cluster stability &
Seeds 0--19; cluster counts 5, 7, and 9; raw, \(\log(1+P)\), and normalized
features; retain top two clusters; report adjusted Rand index, retained set
Jaccard, and rank Spearman correlation. &
Descriptive stability outputs are present for all four v2a-RSNs recordings. \\
Matched baselines &
Raw; rank 40 PCA; rank 40 seeded random orthogonal subspace; energy matched
PCA; one sided causal low pass filters at 0.1, 0.2, and 0.4~Hz. &
Matched controls often exceed the best scoring BSS variant in the current summaries. \\
Simple behavior decoder &
Lags \(\{0,1,2\}\); target shift 0; ridge \(\alpha=10\); training fold bout
quantile 0.75; within variant and raw to variant transfer. &
Used as the practical supervisory readout for larger v2a-RSNs recordings. \\
Null and uncertainty &
Null block sizes 30, 60, and 120; seeds 0--4; non-overlapping 60-frame
uncertainty blocks; 2,000 bootstrap and 5,000 sign permutation replicates. &
Block shuffle and circular shift nulls, plus prediction level uncertainty
outputs, are present; time reversal, phase randomized nulls, BPI intervals,
and causal state intervals are not part of the final reporting. \\
Causal state diagnostic &
Window setting 15, yielding overlapping 14-frame current and shifted
histories; target shifts \(\{0,1,2,4\}\); latent dimension 3; ridge
\(\alpha=10\); linear and shallow tanh MLP transitions; 16 hidden units and
maximum 500 MLP iterations. &
Linear summaries are more interpretable than MLP summaries because several MLP
fits do not converge. \\
Recorded outputs &
Trace and temporal metrics; behavior metrics and predictions; BPI components
and ablations; prediction level uncertainty; causal metrics and out of fold
embeddings; selection and stability; leakage audit; provenance and run
manifest; recording and fish level aggregates. &
Required fold-local outputs are present for \texttt{fish-1}, \texttt{fish-2},
and \texttt{fish-4}; \texttt{fish-3} is omitted from the leakage-audited grid. \\
\hline
\end{tabular}
\end{minipage}
\end{suppwide}
\vspace{0.75\baselineskip}

Table~\ref{tab:supp_v2a_strict_diagnostic_summary} uses the reported
three-recording strict-grid artifacts. It reinforces the main result: BSS can
improve behavior readouts, but matched controls explain much of the gain and
give stronger dynamic state diagnostics.

\clearpage

\begin{suppwide}
\begin{minipage}{\suppdisplaywidth}
\centering
\footnotesize
\renewcommand{\arraystretch}{1.05}
\setlength{\tabcolsep}{4pt}
\captionof{table}{Three-recording strict-grid diagnostic summaries.}
\label{tab:supp_v2a_strict_diagnostic_summary}
\begin{tabular}{@{}p{0.19\linewidth}p{0.29\linewidth}p{0.40\linewidth}@{}}
\hline
\shortstack{Diagnostic} & \shortstack{Statistic summarized} & \shortstack{Across-recording summary} \\
\hline
Reported recordings &
Completion status for the leakage-audited grid. &
\texttt{fish-1}, \texttt{fish-2}, and \texttt{fish-4} are complete.
\texttt{fish-3} is omitted from this grid because the full fold-local run is
computationally too expensive for its 412-variable state space. \\
Behavior readouts &
Best raw, BSS, and matched-control scores in the reported grid. &
For \texttt{fish-1}, vigor \(r\) is 0.594, 0.607, and 0.624 for raw, BSS, and
control; bout BA is 0.736, 0.761, and 0.770. For \texttt{fish-2}, vigor \(r\)
is 0.033, 0.149, and 0.108; bout BA is 0.500, 0.560, and 0.533. For
\texttt{fish-4}, vigor \(r\) is 0.247, 0.341, and 0.354; bout BA is 0.602,
0.624, and 0.641. \\
Dynamic MSE &
Best fold-mean normalized one-step latent transition MSE; lower is better. &
The matched control has the lowest dynamic MSE in all three reported recordings
(0.0001, 0.0018, and 0.0016), compared with BSS values of 0.0072, 0.0027, and
0.0068. \\
Artefact probe &
Residual artefact claim boundary. &
Pseudo-artefact probe rows are present in the compact export, but residual
artefact \(Z\) scores and forward--reverse gap analyses remain absent, so no
residual artefact claim is made. \\
\hline
\end{tabular}
\end{minipage}
\end{suppwide}
\vspace{0.75\baselineskip}

\subsection{Whole recording v2a-RSNs cluster progression}
\label{sec:supp_v2a_cluster_progression}

Tables~\ref{tab:supp_v2a_cluster_220119} and
\ref{tab:supp_v2a_cluster_220127} give the first two descriptive
whole-recording tables behind the BPI diagnostics in the main article.
These tables are transductive because decomposition and cluster selection used
the complete recording. BPI is normalized to raw traces, which are 100 by
construction. Vigor \(r\) is within representation tail vigor Pearson
correlation, bout BA is within representation bout balanced accuracy, trace
\(r\) is median per neuron correlation with raw traces, and NRMSE is mean
per neuron normalized RMSE. The paired tables show the cross recording
heterogeneity clearly: \texttt{fish-1} stays close to raw for Infomax and SOBI,
whereas \texttt{fish-2} shows large intermediate-\(k\) BPI peaks.

\begin{suppwide}
\centering
\begin{minipage}[t]{0.49\suppdisplaywidth}
\centering
\scriptsize
\renewcommand{\arraystretch}{0.96}
\setlength{\tabcolsep}{1.9pt}
\captionof{table}{\texttt{fish-1} Cluster progression; highest BPI marks best \(k\).}
\label{tab:supp_v2a_cluster_220119}
\begin{tabular}{lrrrrrr}
\hline
\shortstack{Method} & k & BPI & \shortstack{Vigor\\\(r\)} & \shortstack{Bout\\BA} & \shortstack{Trace\\\(r\)} & NRMSE \\
\hline
FastICA & 1 & 51.1 & 0.322 & 0.602 & 0.523 & 0.836 \\
 & 2 & 67.6 & 0.445 & 0.641 & 0.619 & 0.761 \\
 & 3 & 73.6 & 0.478 & 0.669 & 0.706 & 0.686 \\
 & 4 & 80.9 & 0.507 & 0.694 & 0.796 & 0.588 \\
 & 5 & 89.6 & 0.546 & 0.710 & 0.893 & 0.446 \\
 & 6 & 90.1 & 0.547 & 0.709 & 0.924 & 0.370 \\
\hline
Infomax & 1 & 100.5 & 0.583 & 0.748 & 0.622 & 0.763 \\
 & 2 & 102.1 & 0.592 & 0.750 & 0.645 & 0.733 \\
 & 3 & 102.1 & 0.594 & 0.752 & 0.704 & 0.685 \\
 & 4 & 102.5 & 0.605 & 0.750 & 0.708 & 0.673 \\
 & 5 & 102.9 & 0.612 & 0.745 & 0.847 & 0.506 \\
 & 6 & 101.9 & 0.597 & 0.742 & 0.940 & 0.362 \\
\hline
SOBI & 1 & 95.4 & 0.555 & 0.732 & 0.662 & 0.730 \\
 & 2 & 104.0 & 0.611 & 0.753 & 0.719 & 0.662 \\
 & 3 & 103.1 & 0.605 & 0.747 & 0.781 & 0.587 \\
 & 4 & 102.6 & 0.598 & 0.746 & 0.854 & 0.509 \\
 & 5 & 101.5 & 0.600 & 0.742 & 0.917 & 0.397 \\
 & 6 & 99.4 & 0.595 & 0.729 & 0.958 & 0.294 \\
\hline
JADE & 1 & 43.1 & 0.279 & 0.572 & 0.374 & 0.906 \\
 & 2 & 53.5 & 0.351 & 0.605 & 0.558 & 0.809 \\
 & 3 & 73.2 & 0.458 & 0.667 & 0.664 & 0.736 \\
 & 4 & 87.2 & 0.512 & 0.709 & 0.774 & 0.624 \\
 & 5 & 90.1 & 0.537 & 0.712 & 0.883 & 0.448 \\
 & 6 & 95.7 & 0.568 & 0.724 & 0.969 & 0.253 \\
\hline
\end{tabular}
\end{minipage}\hfill
\begin{minipage}[t]{0.49\suppdisplaywidth}
\centering
\scriptsize
\renewcommand{\arraystretch}{0.96}
\setlength{\tabcolsep}{1.9pt}
\captionof{table}{\texttt{fish-2} Cluster progression; highest BPI marks best \(k\).}
\label{tab:supp_v2a_cluster_220127}
\begin{tabular}{lrrrrrr}
\hline
\shortstack{Method} & k & BPI & \shortstack{Vigor\\\(r\)} & \shortstack{Bout\\BA} & \shortstack{Trace\\\(r\)} & NRMSE \\
\hline
FastICA & 1 & 144.1 & 0.020 & 0.505 & 0.455 & 0.865 \\
 & 2 & 166.3 & 0.009 & 0.499 & 0.644 & 0.738 \\
 & 3 & 302.1 & 0.043 & 0.518 & 0.844 & 0.551 \\
 & 4 & 149.1 & 0.008 & 0.505 & 0.882 & 0.487 \\
 & 5 & 240.6 & 0.020 & 0.511 & 0.938 & 0.360 \\
 & 6 & 227.3 & 0.051 & 0.507 & 0.971 & 0.259 \\
\hline
Infomax & 1 & 348.6 & 0.104 & 0.525 & 0.808 & 0.572 \\
 & 2 & 418.4 & 0.113 & 0.534 & 0.847 & 0.536 \\
 & 3 & 375.5 & 0.093 & 0.532 & 0.871 & 0.492 \\
 & 4 & 330.9 & 0.089 & 0.528 & 0.890 & 0.458 \\
 & 5 & 309.6 & 0.082 & 0.519 & 0.933 & 0.368 \\
 & 6 & 157.5 & 0.050 & 0.502 & 0.989 & 0.210 \\
\hline
SOBI & 1 & 217.4 & 0.071 & 0.512 & 0.832 & 0.554 \\
 & 2 & 241.6 & 0.076 & 0.510 & 0.882 & 0.486 \\
 & 3 & 277.9 & 0.083 & 0.513 & 0.924 & 0.396 \\
 & 4 & 309.6 & 0.065 & 0.524 & 0.945 & 0.339 \\
 & 5 & 245.3 & 0.062 & 0.514 & 0.966 & 0.272 \\
 & 6 & 119.8 & 0.039 & 0.498 & 0.994 & 0.126 \\
\hline
JADE & 1 & 245.7 & 0.027 & 0.514 & 0.460 & 0.863 \\
 & 2 & 327.5 & 0.038 & 0.522 & 0.639 & 0.741 \\
 & 3 & 306.6 & 0.045 & 0.516 & 0.794 & 0.614 \\
 & 4 & 219.5 & 0.045 & 0.509 & 0.876 & 0.486 \\
 & 5 & 127.7 & 0.016 & 0.505 & 0.924 & 0.381 \\
 & 6 & 171.9 & 0.032 & 0.504 & 0.976 & 0.224 \\
\hline
\end{tabular}
\end{minipage}
\end{suppwide}
\vspace{0.75\baselineskip}

Tables~\ref{tab:supp_v2a_cluster_220210f1} and
\ref{tab:supp_v2a_cluster_220210f2} give the same per cluster values for the
last two recordings. In \texttt{fish-3}, bout balanced accuracy is not finite,
and the highest BPI appears at early Infomax and SOBI cluster counts even
though trace similarity generally increases with more retained clusters. In
\texttt{fish-4}, Infomax again peaks early, while SOBI reaches raw level
behavior only after six retained clusters. These two tables reinforce the main
result that trace preservation and behavior preservation do not move in
lockstep.

\begin{suppwide}
\centering
\begin{minipage}[t]{0.49\suppdisplaywidth}
\centering
\scriptsize
\renewcommand{\arraystretch}{0.96}
\setlength{\tabcolsep}{1.9pt}
\captionof{table}{\texttt{fish-3} Cluster progression; highest BPI marks best \(k\).}
\label{tab:supp_v2a_cluster_220210f1}
\begin{tabular}{lrrrrrr}
\hline
\shortstack{Method} & k & BPI & \shortstack{Vigor\\\(r\)} & \shortstack{Bout\\BA} & \shortstack{Trace\\\(r\)} & NRMSE \\
\hline
FastICA & 1 & 90.6 & 0.341 & NA & 0.761 & 0.663 \\
 & 2 & 90.7 & 0.337 & NA & 0.776 & 0.636 \\
 & 3 & 94.0 & 0.310 & NA & 0.911 & 0.457 \\
 & 4 & 95.9 & 0.317 & NA & 0.929 & 0.405 \\
 & 5 & 96.9 & 0.342 & NA & 0.966 & 0.291 \\
 & 6 & 97.1 & 0.347 & NA & 0.981 & 0.214 \\
\hline
Infomax & 1 & 128.2 & 0.502 & NA & 0.845 & 0.558 \\
 & 2 & 115.9 & 0.450 & NA & 0.923 & 0.433 \\
 & 3 & 108.4 & 0.410 & NA & 0.940 & 0.394 \\
 & 4 & 106.1 & 0.400 & NA & 0.952 & 0.361 \\
 & 5 & 96.2 & 0.334 & NA & 0.980 & 0.259 \\
 & 6 & 92.8 & 0.312 & NA & 0.989 & 0.190 \\
\hline
SOBI & 1 & 116.0 & 0.448 & NA & 0.917 & 0.442 \\
 & 2 & 105.0 & 0.384 & NA & 0.944 & 0.378 \\
 & 3 & 107.7 & 0.402 & NA & 0.954 & 0.342 \\
 & 4 & 106.1 & 0.393 & NA & 0.965 & 0.302 \\
 & 5 & 103.2 & 0.381 & NA & 0.977 & 0.244 \\
 & 6 & 100.6 & 0.367 & NA & 0.989 & 0.165 \\
\hline
JADE & 1 & 31.8 & -0.012 & NA & 0.531 & 0.838 \\
 & 2 & 58.2 & 0.118 & NA & 0.714 & 0.705 \\
 & 3 & 57.8 & 0.114 & NA & 0.731 & 0.686 \\
 & 4 & 68.4 & 0.173 & NA & 0.807 & 0.597 \\
 & 5 & 83.3 & 0.268 & NA & 0.900 & 0.457 \\
 & 6 & 97.9 & 0.349 & NA & 0.960 & 0.284 \\
\hline
\end{tabular}
\end{minipage}\hfill
\begin{minipage}[t]{0.49\suppdisplaywidth}
\centering
\scriptsize
\renewcommand{\arraystretch}{0.96}
\setlength{\tabcolsep}{1.9pt}
\captionof{table}{\texttt{fish-4} Cluster progression; highest BPI marks best \(k\).}
\label{tab:supp_v2a_cluster_220210f2}
\begin{tabular}{lrrrrrr}
\hline
\shortstack{Method} & k & BPI & \shortstack{Vigor\\\(r\)} & \shortstack{Bout\\BA} & \shortstack{Trace\\\(r\)} & NRMSE \\
\hline
FastICA & 1 & 25.6 & 0.081 & 0.538 & 0.466 & 0.867 \\
 & 2 & 44.3 & 0.137 & 0.541 & 0.652 & 0.754 \\
 & 3 & 55.0 & 0.158 & 0.552 & 0.785 & 0.623 \\
 & 4 & 53.8 & 0.119 & 0.545 & 0.836 & 0.555 \\
 & 5 & 73.9 & 0.153 & 0.582 & 0.877 & 0.491 \\
 & 6 & 94.0 & 0.229 & 0.596 & 0.944 & 0.338 \\
\hline
Infomax & 1 & 121.4 & 0.341 & 0.627 & 0.796 & 0.590 \\
 & 2 & 123.6 & 0.342 & 0.627 & 0.848 & 0.528 \\
 & 3 & 117.5 & 0.316 & 0.623 & 0.893 & 0.468 \\
 & 4 & 112.8 & 0.304 & 0.620 & 0.913 & 0.430 \\
 & 5 & 110.4 & 0.280 & 0.618 & 0.955 & 0.346 \\
 & 6 & 104.5 & 0.265 & 0.610 & 0.973 & 0.282 \\
\hline
SOBI & 1 & 61.6 & 0.176 & 0.561 & 0.765 & 0.610 \\
 & 2 & 88.2 & 0.215 & 0.597 & 0.882 & 0.483 \\
 & 3 & 89.9 & 0.234 & 0.592 & 0.923 & 0.405 \\
 & 4 & 86.7 & 0.226 & 0.582 & 0.945 & 0.338 \\
 & 5 & 98.9 & 0.256 & 0.598 & 0.967 & 0.270 \\
 & 6 & 102.2 & 0.258 & 0.604 & 0.981 & 0.208 \\
\hline
JADE & 1 & 28.7 & 0.102 & 0.534 & 0.363 & 0.919 \\
 & 2 & 44.0 & 0.094 & 0.546 & 0.596 & 0.801 \\
 & 3 & 62.6 & 0.163 & 0.561 & 0.741 & 0.668 \\
 & 4 & 74.1 & 0.190 & 0.570 & 0.858 & 0.523 \\
 & 5 & 79.6 & 0.195 & 0.578 & 0.911 & 0.428 \\
 & 6 & 92.1 & 0.248 & 0.588 & 0.956 & 0.306 \\
\hline
\end{tabular}
\end{minipage}
\end{suppwide}
\vspace{0.75\baselineskip}

\subsection{v2a-RSNs CSL matrix plots}
\label{sec:supp_v2a_csl_artifacts}

Figures~\ref{fig:supp_v2a_csl_fish1}--\ref{fig:supp_v2a_csl_fish4}
show the v2a-RSNs c-GC family connectivity matrices behind
the main v2a-RSN CSL density table. Each grid compares the raw traces with the
FastICA, Infomax, SOBI, and JADE \texttt{cluster\_keep\_top\_04}
reconstructions. Rows are source neurons and columns are target neurons. The
diagonal self entries are visible in the plots, but the edge counts in the
panel titles and the densities in the main v2a-RSN CSL density table use only
off diagonal directed edges. Only c-GC and c-GC* are shown.

\begin{suppwide}
    \centering
    \begin{minipage}[t]{\suppdisplaywidth}
        \centering
        \includegraphics[width=0.48\linewidth]{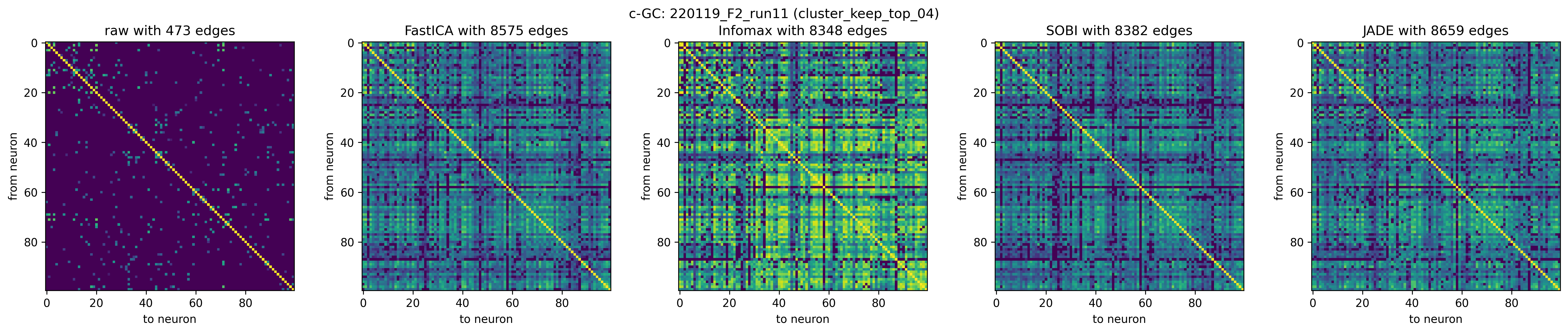}
        \hfill
        \includegraphics[width=0.48\linewidth]{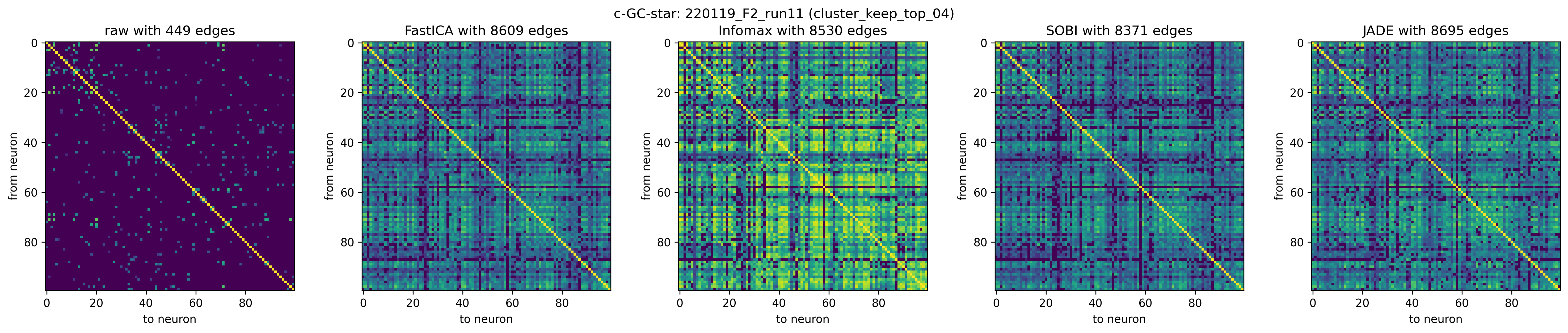}
        \captionof{figure}{v2a-RSN CSL matrices for \texttt{fish-1}: c-GC left, c-GC* right.}
        \label{fig:supp_v2a_csl_fish1}
    \end{minipage}
    \par\vspace{0.1em}
    \begin{minipage}[t]{\suppdisplaywidth}
        \centering
        \includegraphics[width=0.48\linewidth]{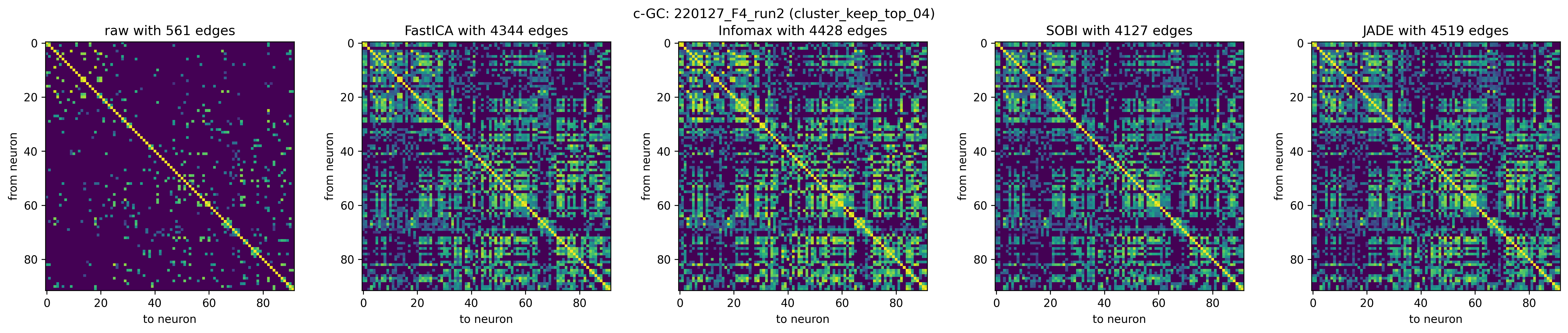}
        \hfill
        \includegraphics[width=0.48\linewidth]{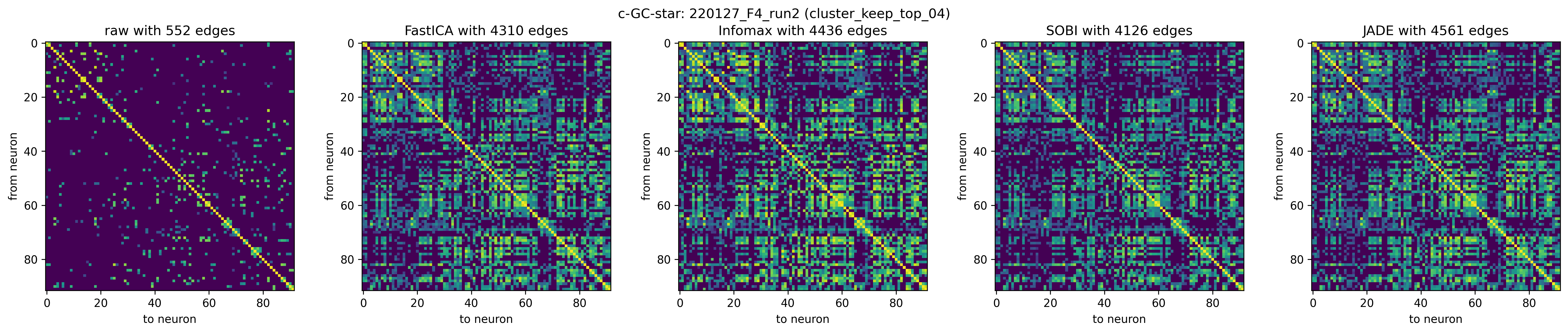}
        \captionof{figure}{v2a-RSN CSL matrices for \texttt{fish-2}: c-GC left, c-GC* right.}
        \label{fig:supp_v2a_csl_fish2}
    \end{minipage}
    \par\vspace{0.1em}
    \begin{minipage}[t]{\suppdisplaywidth}
        \centering
        \includegraphics[width=0.48\linewidth]{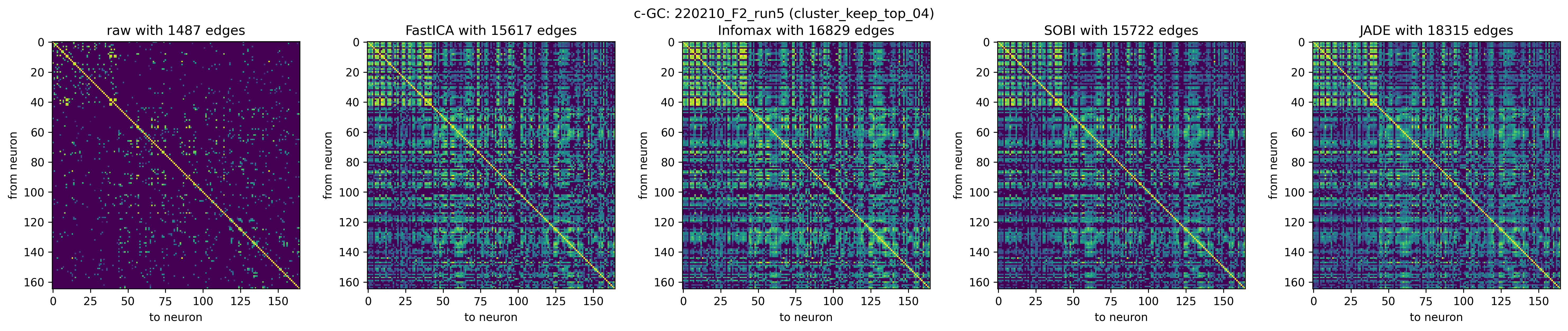}
        \hfill
        \includegraphics[width=0.48\linewidth]{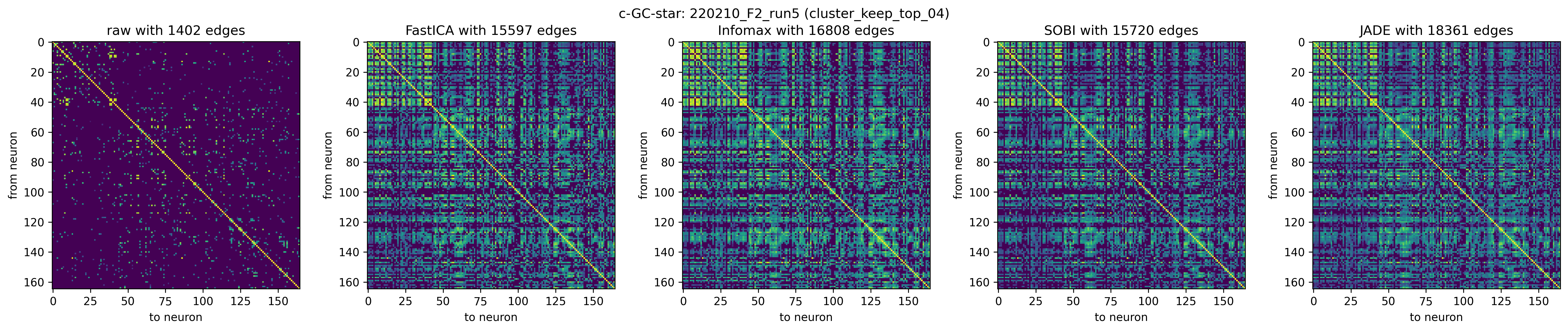}
        \captionof{figure}{v2a-RSN CSL matrices for \texttt{fish-4}: c-GC left, c-GC* right.}
        \label{fig:supp_v2a_csl_fish4}
    \end{minipage}
\end{suppwide}

\subsection{Detailed method definitions}

\paragraph{Joint diagonalization routine.}
\label{sec:supp_joint_diagonalization}
SOBI/JADE share a Jacobi style symmetric joint diagonalizer. For each index
pair, the routine estimates a Givens rotation from off-diagonal and
diagonal-difference terms across the covariance or cumulant matrices, applies
it to all matrices, and accumulates the total rotation. This follows classical
Jacobi simultaneous diagonalization \cite{CardosoSouloumiac1996Jacobi}.

\paragraph{IC feature extraction and scoring.}
\label{sec:supp_ic_quality}
The IC quality table stores one row per component. Temporal features flag
bursty, heavy-tailed or nonstationary time courses; Welch spectra provide power
ratios, entropy and narrowband dominance \cite{Welch1967PSD}; loading features
include norms, maximum-to-median loading ratio, strong-loading fraction and
Hoyer sparsity \cite{Hoyer2004Sparsity}. Optional ROI coordinates give spatial
centroid and extent. Aligned behavior correlations are protective features
because unusual components can still contain behavior-related neural signal.
Artefact and protection scores average percentile-ranked feature evidence.
Components are recommended as \texttt{drop} only when artefact evidence exceeds
the drop threshold and protection evidence remains below its maximum threshold;
otherwise they are marked \texttt{keep} or \texttt{review}.

\paragraph{Component removal impact.}
\label{sec:supp_reconstruction_impact}
For each component or component group, the validation table records the
reconstruction change after removal. Reported quantities include mean absolute
and root mean square perturbation, global trace correlation change, median
per-neuron variance ratio and behavior-correlation changes when aligned targets
are available.

\paragraph{Behavior target construction.}
\label{sec:supp_behavior_targets}
High-rate tail angle samples are binned to calcium frames. Tail vigor is
frame-bin RMS first-difference velocity, smoothed with a centered three-frame
mean before fold construction. Bout state applies a training-fold vigor-quantile
threshold to held-out frames.

% \subsection{Transductive v2a trace progression across BSS methods}

\end{document}